\pdfoutput=1

\documentclass[superscriptaddress,amsmath,amssymb,twocolumn]{revtex4-1}

\usepackage{graphicx}
\usepackage{chemarr}
\usepackage{times,longtable}
\usepackage{hyperref}
\usepackage{color}
\usepackage{afterpage}
\usepackage{soul,comment}
\usepackage[usenames,dvipsnames]{xcolor}
\usepackage{colortbl}
\usepackage{pbox}
\usepackage{capt-of}
\usepackage{pdfpages}
\makeatletter
\AtBeginDocument{\let\LS@rot\@undefined}
\makeatother

\graphicspath{{large/}}

\bibpunct{[}{]}{,}{n}{}{,}

\newcommand{\red}[1]{\textcolor{black}{ #1}}

\newcommand{\ovl}[1]{\overline{#1}}
\renewcommand{\b}{\mathrm{tr}}
\newcommand{\CV}{\mathrm{CV}}
\newcommand{\beq}{\begin{equation}}
\newcommand{\eeq}{\end{equation}}

\newcommand{\avg}[1]{\left\langle #1 \right\rangle}

\usepackage{chemarr}
\usepackage{epsf,mathtools}
\usepackage{graphicx}
\usepackage{dcolumn}
\usepackage{bm}

\begin{document}

\title{\textsf{Competing endogenous RNA crosstalk at system level}}

\author{Mattia Miotto}
\affiliation{Dipartimento di Fisica, Sapienza Universit\`a di Roma, Rome, Italy}

\author{Enzo Marinari}
\affiliation{Dipartimento di Fisica, Sapienza Universit\`a di Roma, Rome, Italy}
\affiliation{INFN, Sezione di Roma 1, Rome, Italy}

\author{Andrea De Martino}
\email{andrea.demartino@roma1.infn.it}
\affiliation{Soft \& Living Matter Lab, Institute of Nanotechnology (CNR-NANOTEC), Consiglio Nazionale delle Ricerche, Rome, Italy}
\affiliation{Italian Institute for Genomic Medicine, Turin, Italy}

\begin{abstract}
\noindent \textbf{Abstract --} microRNAs (miRNAs) regulate gene expression at post-transcriptional level by repressing target RNA molecules. Competition to bind miRNAs tends in turn to correlate their targets, establishing effective RNA-RNA interactions that can influence expression levels, buffer fluctuations and promote signal propagation. Such a  potential has been characterized mathematically for small motifs both at steady state and \red{away from stationarity}. Experimental evidence, on the other hand, suggests that competing endogenous RNA (ceRNA) crosstalk is rather weak. \red{Extended miRNA-RNA networks could however favour the integration of many crosstalk interactions, leading to significant large-scale effects in spite of the weakness of individual links.} To clarify the extent to which crosstalk is sustained by the miRNA interactome, we have studied its emergent systemic features {\it in silico} in large-scale miRNA-RNA network reconstructions. \red{We show that, although generically weak, system-level crosstalk patterns} (i) are enhanced by transcriptional heterogeneities, (ii) can achieve high-intensity even for RNAs that are not co-regulated, (iii) are robust to \red{variability} in transcription rates, and (iv) are significantly non-local, i.e. correlate weakly with miRNA-RNA interaction parameters. \red{Furthermore, RNA levels are generically more stable when crosstalk is strongest. As some of these features appear to be encoded in the network's topology, crosstalk may functionally be favoured by natural selection.} These results suggest that, besides their repressive role, miRNAs mediate a weak but resilient and context-independent network of cross-regulatory interactions that interconnect the transcriptome, stabilize expression levels and support system-level responses. \\
~\\
\textbf{Author summary --} Large regulatory networks integrate a huge of number molecular interactions into robust system-level outcomes. This capability can emerge even when individual interactions are weak and/or strongly heterogeneous. We show this in the context of human post-transcriptional regulation driven by microRNAs (miRNAs). These small non-coding RNAs mediate an extended network of weak cross-regulatory interactions between their targets. We characterize such a network {\it in silico} using a variety of quantitative measures. Despite their weakness, miRNA-mediated couplings constitute a highly interconnected regulatory layer with robust interaction patterns that contribute to the stabilization of expression levels and allow for tunable system-level responses to specific signals. As some of these features are encoded, to a large degree, in the network's topology, natural selection appears to have favored the evolution of this ``soft mode'' of cross-regulation between RNAs.
\end{abstract}

\maketitle

\section*{Introduction}

Competition to bind substrates, enzymes or gene expression machinery is ubiquitous in biological networks and impacts regulatory processes in several ways \cite{grigorova,buchler,cookson,chu,devos,brackley,rondelez,cotari,mather,brewster,gyorgy,mauri}. For instance, the initiation and translation rates of different transcripts are effectively coupled by the competition for the ribosome pool, \red{so that modifications of a given RNA species can alter the translational dynamics of other RNAs} \cite{raveh}. Quite generally, competition for limited and shared molecular resources induces effective interactions between the competing species, with signs (positive or negative) that depend on the specifics of the underlying processes \cite{wei}. \red{While such interactions constitute in principle an additional layer of indirect regulation, their intensity is  strongly context-dependent} \cite{mather}. The \red{functional} role of competition-driven crosstalk therefore has to be evaluated on a case-by-case basis. 

Competition for miRNAs (or, more generally, small regulatory RNAs) among long transcripts is undergoing much scrutiny \red{in this respect} \cite{salmena}. {\it In silico} studies of small motifs, summarized e.g. in \cite{lai,martirosyan}, have characterized how the strength, selectivity and directionality of miRNA-mediated RNA crosstalk are modulated by kinetic and topologic ingredients, leading to highly adjustable output profiles \cite{figliuzzi,bosia,tian}, differential processing of intrinsic and extrinsic heterogeneities \cite{mehta,martirosyan2,re,delgiudice}, stabilization of protein levels \cite{schmiedel,martirosyan3} and long-range effects \cite{nitzan}, both at steady state and during transients \cite{figliuzzi2}. Experimental evidence, however, suggests that, in order to fully develop its potential, RNA crosstalk presupposes rather specific conditions, either in terms of the size of the perturbation required to generate a significant response \cite{tay,denzler,denzler2} or in terms of molecular abundances and kinetic parameters \cite{bosson,yuan,bosia2} (see e.g. \cite{jens,thomson} for reviews). \red{When such conditions are not met, miRNAs only provide a weak coupling channel for RNAs.}

Generally speaking, weak individual crosstalk interactions by themselves do not necessarily imply a reduced physiological role. \red{This is especially true in large networks}, where many interactions can aggregate and perturbations can propagate by exploiting topological and kinetic heterogeneities \cite{figliuzzi,martirosyan2,chiu}. On-going explorations of miRNA-RNA networks are indeed uncovering a high degree of hard-wired complexity \cite{rzepiela,mcgeary}. In the light of these studies, achieving a better understanding of RNA crosstalk from a systemic perspective has become a pressing issue. 

\red{Our goal here is to examine RNA crosstalk {\it in silico} in extended miRNA-RNA interactomes as a function of various parameters, including global miRNA levels, degrees of parameter heterogeneity, and topological characteristics of the networks. To cope with the lack of knowledge about kinetic parameters, we make use of a maximum-entropy assumption \cite{jaynes}. In short, after obtaining the steady states of the miRNA-RNA network in terms of a small number of kinetic parameters, we focus on the statistics of different quantities induced by a probability distribution of these parameters. This allows to extract context-independent, or typical, features at the cost of weakening our ability to make predictions for individual crosstalk interactions.}

In short, our main results can be summarized as follows. 
\begin{enumerate}
\item Although typically weak, the emergent crosstalk structure is a robust feature of the miRNA-RNA network; for instance, its mean intensity is modulated by miRNA levels but is otherwise weakly affected by transcriptional and/or kinetic heterogeneities (including binding affinities).
\item On the other hand, variability in transcription rates generically enhances the maximal crosstalk intensity achievable as well as non-local effects (i.e. the emergence of long-range  crosstalk mediated by chains of miRNA-RNA interactions).
\item The stability of expression profiles is generically higher when crosstalk is strongest.
\item The degrees of RNA and miRNA nodes are the key topological controllers of the above picture.
\end{enumerate}
Overall, these points suggest that miRNA-RNA networks \red{encode for complex and adaptive crosstalk patterns} that feed back on the stability of expression profiles despite the fact that the typical crosstalk link is very weak. A relatively small number of stronger couplings drives this scenario, \red{while transcriptional and topologic heterogeneities allow to extend the range of crosstalk up to network scale.}

\section*{Results}

\subsection*{Mathematical model}

\begin{figure*}
\begin{center}
\includegraphics[width=\textwidth]{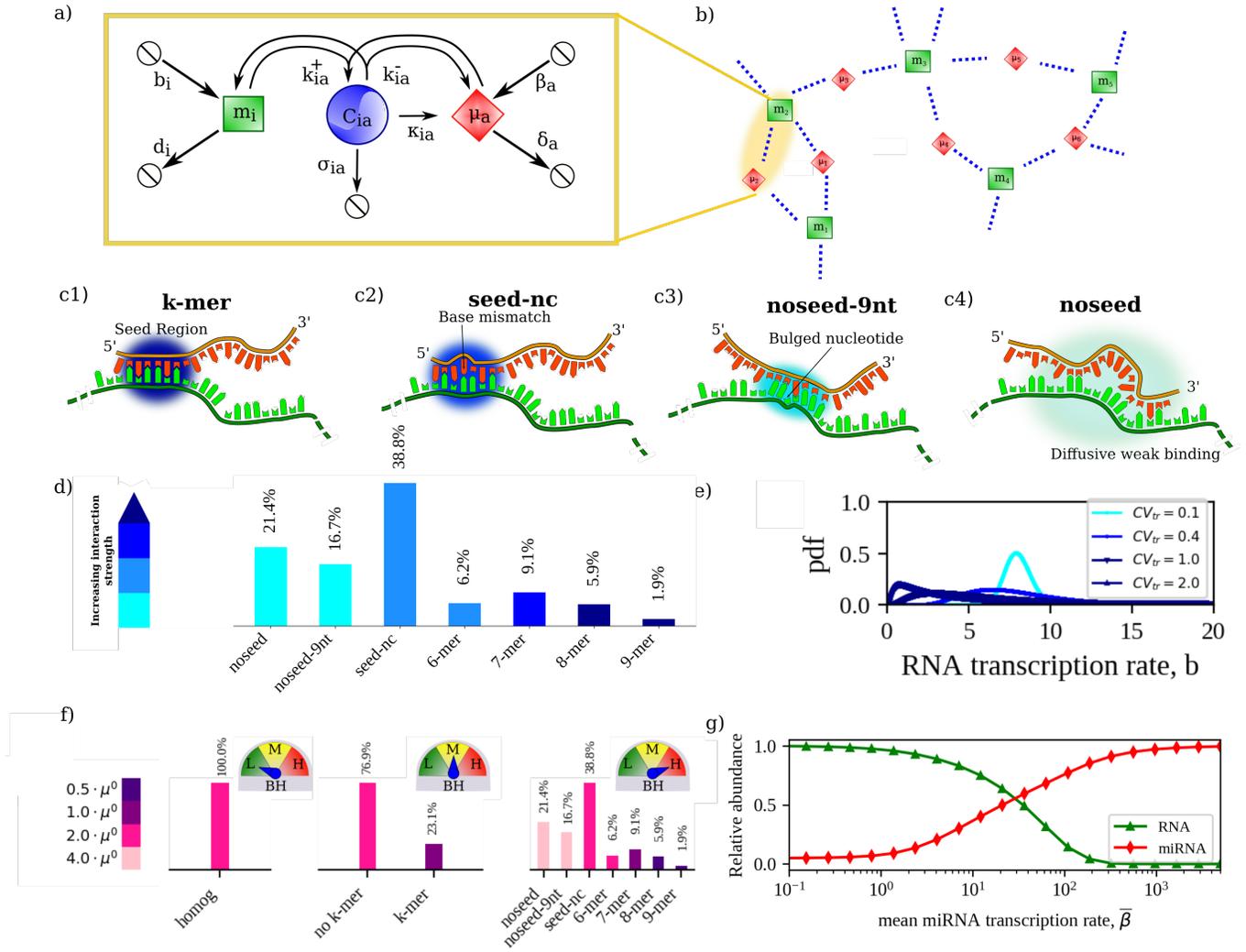}
\end{center}
\caption{{\bf Basic modeling and network features.} \textbf{(a)} Scheme of the interaction between RNA species $i$ and miRNA species $a$. Synthesis, degradation, association and dissociation processes are shown with their respective rates. \textbf{(b)} Sketch of a bipartite miRNA-RNA network. RNA species can crosstalk via chains of miRNA-mediated effective interactions, as can do species 1 and 3 in this example. \textbf{(c1-c4)} Classes of miRNA-RNA interactions considered in this work (following \cite{helwak}): (c1) perfect k-base-pairing in the seed region ('k-mer' mode); (c2) seed base-pairing with up to one mismatched or bulged nucleotide (non-canonical mode or `seed-nc'); (c3) non-seed base pairing with bulged and/or mismatched nucleotides ('noseed-9nt' mode); (c4) non-seed binding within weak diffuse regions ('noseed' mode). \textbf{(d)} Frequency of each binding mode in the CLASH dataset (from \cite{helwak}). \textbf{(e)} Distributions of RNA transcription rates used in this work: each rate is assumed to be drawn independently from a lognormal distribution with given mean (same for each RNA species). Increased transcriptional heterogeneity (TH) corresponds to increased values of the relative fluctuations ($\CV_\b$). \textbf{(f)} Scenarios of miRNA-RNA binding heterogeneity (BH) analyzed in this work. From left to right: low BH, where each miRNA-RNA pair interacts with the same strength; medium BH, with k-mer interactions (stronger) distinguished from the rest (weaker); high BH, where the full structure described in panel (d) is employed. \textbf{(g)} Relative abundance of miRNAs  and RNAs as a function of the global average miRNA transcription rate in the CLASH interactome for representative values of kinetic parameters.  
The `susceptible' regime in which global RNA levels are more sensitive to variations in miRNA availability occurs at intermediate values of $\ovl{\beta}$.}
\label{fig1}
\end{figure*}

To model a network comprising $M$ miRNA species and $N$ RNA species we have extended the mathematical framework employed in \cite{figliuzzi} for the study of small motifs. Conforming to experimental evidence according to which mature miRNAs are mostly bound to Argonaute \cite{burroughs}, the model assumes molecule availability as the only inhibition-limiting factor and describes the interaction between miRNA species $a$ (ranging from 1 to $M$) and RNA species $i$ (ranging from 1 to $N$) in terms of (see Fig\,\ref{fig1}a)
\begin{itemize}
\item synthesis rates ($b_i$ for the RNA, $\beta_a$ for the miRNA)
\item degradation rates ($d_i$ for the RNA, $\delta_a$ for the miRNA)
\item miRNA-RNA association and dissociation rates ($k_{ia}^+$ and $k_{ia}^-$ respectively)
\item miRNA-RNA complex degradation rates ($\kappa_{ia}$ for the catalytic pathway leading to miRNA recycling \cite{baccarini} and $\sigma_{ia}$ for the stoichiometric pathway without recycling)
\end{itemize}
Assuming deterministic mass-action kinetics, molecular levels ($m_i$ for RNA species $i$, $\mu_a$ for miRNA species $a$) evolve according to 
\begin{eqnarray}
\label{eq:sup:timeevol}
\nonumber\frac{d \mu_a}{dt} &=& \beta_a - \delta_a \,\mu_a - \sum_{i=1}^N k_{ia}^+ m_i\, \mu_a + \sum_{i=1}^N (k_{ia}^-  +  \kappa_{ia}) c_{ia}~~,\\
\frac{dm_i}{dt} &=& b_i - d_i \,m_i - \sum_{a=1}^M \left(k_{ia}^+m_i \, \mu_a -  k_{ia}^- c_{ia}\right),\\
\nonumber\frac{d c_{ia}}{dt} &=&  k_{ia}^+ m_i\,\mu_a - ( k_{ia}^-  + \kappa_{ia} + \sigma_{ia}) c_{ia}~~,
\end{eqnarray}
where $c_{ia}$ denotes the level of the complexes formed by RNA species $i$ and miRNA species $a$. Such a system possesses a unique asymptotically stable steady state \cite{flondor}, where molecular levels attain the values (here and in the following, we denote the steady state value of variable $x$ as $[x]$)
\begin{gather}
\nonumber[\mu_a]= \frac{\mu_a^\star}{1 + \sum_{i=1}^{N} \frac{[m_i]}{m^0_{ia}}}~~,\label{ssmu}\\
[m_i] = \frac{m_i^\star}{1 +\sum_{a=1}^{M} \frac{[\mu_a]}{\mu^0_{ia}}}~~,\label{ssm}\\
\nonumber[c_{ia}] = \frac{ k_{ia}^+[m_i][\mu_a] }{ ( k_{ia}^-  + \kappa_{ia} + \sigma_{ia}) }~~,\label{ssc}
\end{gather}
with $m_i^\star \equiv b_i/d_i$ and $\mu_a^\star\equiv \beta_a/\delta_a$ the concentrations of \red{RNA and miRNA species} in absence of inhibition. The quantities $\mu^0_{ia}$ and $m^0_{ia}$ are given respectively by
\begin{gather}\label{mu0}
\mu_{ia}^0 =  \frac{d_i}{k_{ia}^+}\left( 1 + \frac{k_{ia}^-}{\sigma_{ia} + \kappa_{ia}}\right)~~,\\
m_{ia}^0 = \frac{\delta_a}{k_{ia}^+}\left( 1 + \frac{k_{ia}^{-}+ \kappa_{ia}}{\sigma_{ia}}\right) \label{m0}~~,
\end{gather}
and effectively quantify the inverse repression strengths of miRNAs and RNAs. Specifically, see Eq (\ref{ssm}), RNA species $i$ is unrepressed (or, respectively, repressed) by miRNA species $a$ when $[\mu_a]\ll\mu^0_{ia}$ ($[\mu_a]\gg\mu^0_{ia}$). Hence the smaller is $\mu^0_{ia}$ the stronger the repression that $a$ can exert on $i$. Similar considerations hold for $m^0_{ia}$: the smaller it is, the more miRNA species $a$ will be sequestered by RNA species $i$. 

The strength of miRNA-mediated RNA crosstalk at steady state can be estimated by the change in the steady-state level of RNA species $i$ induced by a (small) variation in the transcription rate of species $j$, quantified by the {\it susceptibility} \cite{figliuzzi}
\begin{equation}
\chi_{ij} \equiv d_j\frac{\partial [m_i]}{\partial b_j}~~.
\label{eq:Chi}
\end{equation} 
\red{(The prefactor $d_j$ in Eq (\ref{eq:Chi}) serves the only purpose of making $\chi_{ij}$ dimensionless.) Note that} (i) $\chi_{ij}\geq 0$ (i.e. the effective interaction tends to increase or decrease the levels of both RNAs) and (ii) $\chi_{ij}$ and $\chi_{ji}$ are {\it a priori} different (see  \cite{martirosyan4} for a detailed discussion of this aspect). The advantage of the susceptibility over alternative measures of crosstalk, like the Pearson correlation coefficient, lies in the fact that it focuses on the role of competition, disregarding indirect effects due e.g. to fluctuations in miRNA levels. An extended comparison of different crosstalk measures can be found in \cite{martirosyan}.

Starting from Eq (\ref{ssm}), \red{one can derive an analytical expression allowing for the convenient computation of $\chi_{ij}$ for each $(i,j)$ pair in any miRNA-RNA network specified by a given set of kinetic parameters} (see \underline{Supporting Text}, Section 1). In compact form, the susceptibility matrix $\boldsymbol{\widehat\chi}=\{\chi_{ij}\}_{i,j=1}^N$ turns out to be given by
\begin{equation}\label{chimatrix}
\boldsymbol{\widehat{\chi}} = \left( \mathbf{\widehat{1}- \widehat{W}} \right)^{-1} \mbox{diag} \left( \mathbf{\frac{m}{m^\star}}\right) ~~,
\end{equation}
where $\mbox{diag} \left( \mathbf{\frac{m}{m^\star}}\right)$ denotes the $N$-dimensional diagonal matrix with elements $m_i/m_i^\star$ ($i=1,\ldots,N$) while $\mathbf{\widehat{W}}$ is an $N\times N$ matrix with elements 
\begin{equation}
\label{eq:W}
W_{ij} = \frac{[m_i]^2}{m_i^\star}\sum_{a \in (i\cap j)}\frac{1}{m_{ja}^0\mu_{ia}^0}\frac{[\mu_a]^2}{\mu_a^\star} ~~,
\end{equation}
the sum running over all miRNA species that co-target RNAs $i$ and $j$. Note that Eq (\ref{chimatrix}) implies that $i$ and $j$ need not be targeted by a common miRNA species in order for $\chi_{ij}$ to be non-zero, as crosstalk can propagate through chains of miRNA-mediated interactions \cite{martirosyan,nitzan} (see Fig \ref{fig1}b). A toy model explicitly displaying this mechanism is discussed in \underline{Supporting Text}, Section 2.

\subsection*{Choice of \red{networks} and parameters, and simulated scenarios}

We shall mainly focus on the human miRNA interactome reconstructed in \cite{helwak} using the CLASH (Crosslinking, Ligation And Sequencing of Hybrids) protocol. We refer to this as the `CLASH interactome' for short; see Materials and Methods for details. \red{The 4 types of miRNA-RNA couplings we consider are described in Fig~\ref{fig1}c: (c1) perfect pairings of $k$ miRNA seed nucleotides (``k-mer'' for brevity, with $k$ ranging from 6 to 9); (c2) sequence-specific pairings with up to one bulge or mismatch in the seed region (non-canonical pairings, or ``seed-nc'' for short); (c3) a 9 nt stems no-seed interaction allowing for bulged nucleotides in the target (``noseed-9nt''); and (c4) a no-seed interaction with distributed weak pairings (``noseed'' for short). Non-canonical pairings are the most abundant in the CLASH interactome, accounting for roughly 77\% of all miRNA-RNA interactions \cite{helwak} (see Fig \ref{fig1}d). They are also weaker than canonical ones and seem to exert a very limited repressive role \cite{agarwal}. Nevertheless, they in principle contribute to miRNA titration and hence to RNA crosstalk.} We therefore included them in our analysis. Our results will however turn out to be qualitatively independent of whether non-canonical sites are accounted for. A summary of basic features of the CLASH subnetworks spanned by different classes of interactions is given in \underline{Supporting Text}, Supplementary Table 1. 

For sakes of simplicity, we assume $\kappa_{ia} = \kappa$ and $\sigma_{ia} = \sigma$ for each $(i,a)$ pair, $\delta_{a} = \delta$ for all $a$ and $d_{i} = d$ for all $i$. With this choice, one has, for each miRNA-RNA pair,
\begin{equation}\label{condit}
\frac{\mu_{ia}^0}{m_{ia}^0} \equiv \frac{d_i}{\delta_a}\,\frac{\sigma_{ia}}{\sigma_{ia}+\kappa_{ia}}=\frac{\lambda d}{\delta}~~,
\end{equation}
with $\lambda = \frac{\sigma}{\sigma + \kappa}$ the `stoichiometricity ratio'. Using values of $\lambda$, $d$ and $\delta$ compatible with empirical evidence (see Table~\ref{tab:data}), we set $\mu_{ia}^0/m_{ia}^0\simeq 0.59$. 
\begin{table}
\caption{Summary of parameter values.}
\label{tab:data}
\begin{tabular}{@{}llll@{}}
\hline
\hline
Parameter & Value  & Description& Ref.
\\
\hline
\hline
$d$& 0.08 [h$^{-1}$]&  RNA degradation rate &\cite{chiu}
\\
$\delta$& 0.027 [h$^{-1}$]&  miRNA degradation rate &\cite{chiu}
\\
$\lambda$& 0.2&  stoichiometricity ratio &\cite{chiu}
\\
$\ovl{b}$& 8 [molecules/h] &  mean RNA transcription rate & \cite{chiu} 
\\
$\mu^{0}$ &  4 [molecules]&&\\
\hline
\hline
\end{tabular}%
\end{table}
With this choice, network parameters are fully determined by specifying (i) transcription rates ($\beta_a$ for miRNAs and $b_i$ for RNAs), and (ii) the values of either $\mu^0_{ia}$ or $m^0_{ia}$ for each miRNA-RNA pair. \red{We shall consider different scenarios for these quantities (see below).} Once parameters are set, emergent crosstalk patterns are obtained by solving Eq (\ref{chimatrix}) numerically. 

\subsubsection*{Transcription rates and transcriptional heterogeneity (TH)} 
 
Throughout this study, we assume that both RNA transcription rates $b_i$ ($i=1,\ldots,N$) and miRNA transcription rates $\beta_a$ ($a$ = 1, ..., M) are log-normal i.i.d. random variables with means $\ovl{b}$ and $\ovl{\beta}$, and variances $\sigma_b^2$ and $\sigma_\beta^2$, respectively. The mean RNA transcription rate $\ovl{b}$ is kept fixed at 8 molecules/h (see Table~\ref{tab:data}), while we use the mean miRNA transcription rate $\ovl{\beta}$ as a control parameter upon varying which crosstalk patterns are analyzed. To assess the impact of heterogeneity in transcription rates across molecular species (TH for short), we study how crosstalk patterns change as the magnitude of fluctuations increases, assuming the same transcriptional variability for miRNAs and RNAs. \red{Our goal is to understand how the effective interaction network processes different degrees of variability in transcription rates, particularly at the level of RNAs. We hence tune TH by changing the coefficient of variation of individual rates (standard deviation over mean, see Fig \ref{fig1}e), which we denote by $\CV_\b$. In particular, we have exploited the log-normality of transcription rates to explore a 20-fold range of values of $\CV_\b$, from $\CV_\b=0.1$ to $2$.}

\begin{figure*}[t!]
\begin{center}
\includegraphics[width=\textwidth]{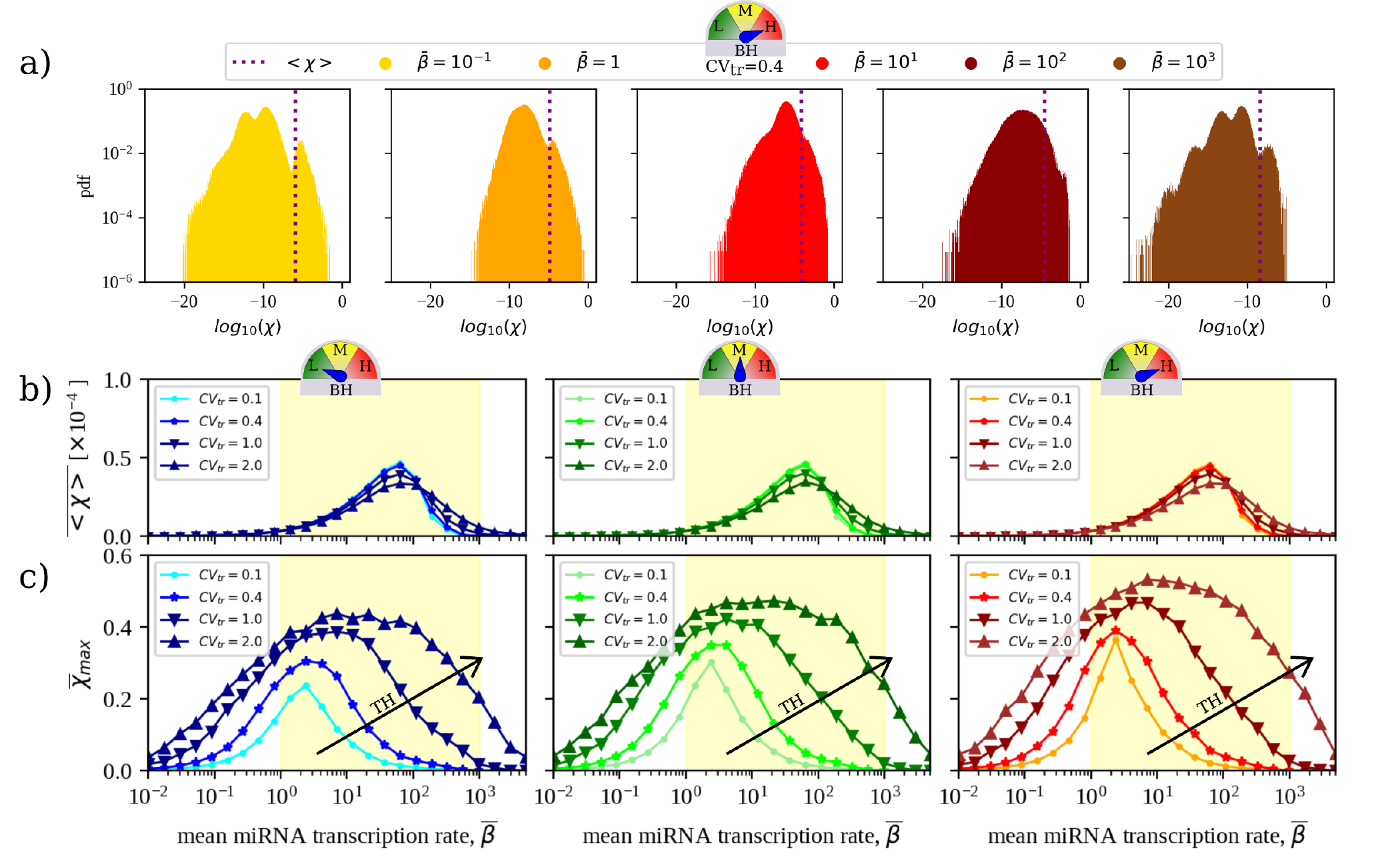}
\end{center}
\caption{\textbf{Quantitative features of RNA crosstalk in the CLASH interactome derived by solving Eq (\ref{chimatrix}).} \textbf{(a)} Representative distributions of susceptibilities obtained for the CLASH interactome for five different realizations of parameters with different values of $\ovl{\beta}$, $\CV_\b=0.4$ and maximal BH. \textbf{(b)} Mean susceptibility (averaged over pairs of distinct RNAs and over 100 independent realizations of TH) as a function of the mean miRNA transcription rate $\ovl{\beta}$. \textbf{(c)} Mean maximal susceptibility (computed over all pairs of distinct RNAs and averaged over 100 independent realizations of TH) as a function of the mean miRNA transcription rate $\ovl{\beta}$. Results are shown for the 3 BH scenarios considered. Parameter values are reported in Table \ref{tab:data}. The yellow shaded area qualitatively marks the region where the mean susceptibility is significantly different from zero, which coincides with the susceptible regime \cite{figliuzzi}. In each case, the standard error of the mean is equal to or smaller than the size of the markers.}
\label{fig2}
\end{figure*}

\subsubsection*{Binding strengths and binding heterogeneity (BH)}

To appraise how heterogeneities in the miRNA-RNA interaction strengths (binding heterogeneity or BH for short) affect the emergent crosstalk landscape, we consider three variants of the structure of binding affinities encoded in the CLASH interactome (see Fig \ref{fig1}f). At the lowest level \red{of diversity}, we assume a homogeneous network in which $\mu^0_{ia}=2\mu^0$ for each miRNA-RNA pair, with $\mu_0$ a constant taken to be equal to 4 molecules (see Table~\ref{tab:data}). (Assuming a negligible miRNA-RNA unbinding rate, this corresponds roughly to an association rate of $0.02/$molecules/hour, in agreement with values reported in \cite{chiu}.) At the intermediate level, we discriminate $(i,a)$ pairs interacting via k-mer pairing (for which we take $\mu^0_{ia}=\mu^0$, i.e. stronger coupling) from the rest ($\mu^0_{ia}=2\mu^0$). Finally, at the highest level we associate different binding strengths to each of the four types of miRNA-RNA pairs, assuming a 2-fold change in $\mu^0_{ia}$ between groups in agreement with estimates from \cite{helwak}. In each case, $m^0_{ia}$'s are computed from Eq (\ref{condit}).

\subsubsection*{Role of network topology}

We furthermore characterize the extent to which crosstalk patterns are induced by the specific wiring of the CLASH interactome by comparing it to the patterns that arise in randomized versions of the same network. In particular, we study ensembles of networks obtained by re-assigning each $(i,a)$ link to a miRNA-RNA pair drawn randomly among all possible pairs with equal probability. This type of re-wiring disregards topological correlations of all orders, including node connectivities \cite{coolen}. To evaluate the impact of the specific degree sequences encoded in the mapped miRNA interactome on the emergent system-level crosstalk patterns, we also analyze networks generated by a more conservative procedure based on degree-preserving edge-swaps \cite{coolen2}. Details of the latter are given in \underline{Supporting Text}.

\subsection*{The mean RNA crosstalk intensity is a robust property of the miRNA-RNA network}

Expectedly, the overall abundance of free RNAs and free miRNAs change in opposite directions as miRNA transcription is globally upregulated and RNAs are increasingly repressed (see Fig \ref{fig1}g). Susceptibilities are bound to be larger when RNAs are more sensitive to changes in miRNA levels, i.e. in the so-called susceptible region at intermediate values of $\ovl{\beta}$ \cite{martirosyan}. \red{Representative susceptibility distributions derived by solving Eq (\ref{chimatrix}) and describing the CLASH network's crosstalk pattern for different degrees of TH and BH are displayed in Fig \ref{fig2}a and in  \underline{Supporting Text}, Supplementary Fig S4. Fig \ref{fig2}b focuses instead on a systemic feature, namely the mean susceptibility $\ovl{\avg{\chi}}$, where the brackets $\avg{\cdots}$ denote an average over all pairs of distinct RNA species while the over-bar stands for an average over different realizations of transcription rate profiles at fixed $\CV_\b$. $\ovl{\avg{\chi}}$ informs about the typical strength of RNA crosstalk in the network and is shown as a function of the mean transcription rate of miRNAs.}

miRNA availability modulates $\ovl{\avg{\chi}}$ so that it peaks within the susceptible region and is vanishingly small outside of it (see Fig \ref{fig2}b), where molecular levels are practically unaffected by varying miRNA transcription rates. Notably, this picture is substantially unchanged by modifying the degrees of TH and/or BH, save for a \red{modest} expansion of the susceptible region. \red{Such a behaviour therefore describes a `basal level' of crosstalk that occurs in the network in any given condition. To appraise its significance, one can gauge it against the self-susceptibility $\chi_{ii}=d_i\frac{\partial m_i}{\partial b_i}$, which quantifies the change in the level of free transcripts of species $i$ induced by a small modification of its own transcription rate. (Note that, by definition, $\chi_{ii}\leq 1$.) \underline{Supporting Text}, Supplementary Fig S5 displays the mean self-susceptibilities computed in the conditions of TH and BH of Fig \ref{fig2}b. One sees that $\ovl{\avg{\chi}}$ is about four orders of magnitude smaller than the mean self-susceptibility. In this respect, crosstalk appears to be on average very weak.}

On the other hand, the picture just derived strongly suggests that the mean susceptibility profile is determined to a large extent by the topology of the network. We shall validate this hypothesis in the following. \red{This conclusion, as well as the overall qualitative crosstalk characteristics illustrated by Fig \ref{fig2}a, will be seen to remain valid also when the contribution of non-canonical binding sites is disregarded.}

\subsection*{The achievable crosstalk strength is enhanced by transcriptional heterogeneities}

\red{Fig \ref{fig2}c displays the behaviour of the mean maximum susceptibility $\ovl{\chi}_{\max}=\ovl{\max_{(i,j)} \chi_{ij}}$, where the maximum is taken over all pairs of different RNA species (i.e. with $i\neq j$). $\ovl{\chi}_{\max}$ quantifies the maximum achievable intensity of crosstalk interactions in each scenario,  therefore providing a proxy for the strength of the most significant miRNA-mediated couplings arising between different RNA species in the network. Like $\ovl{\avg{\chi}}$, $\ovl{\chi}_{\max}$ also peaks in the susceptible regime, albeit for smaller values of the mean miRNA transcription rate. The strongest crosstalk hence typically occurs when RNA levels are just weakly sensitive to changes in miRNA availability. Remarkably, $\ovl{\chi}_{\max}$ is four orders of magnitude larger than $\ovl{\avg{\chi}}$. The backbone of the RNA crosstalk network formed by the most intense interactions is therefore comparable in intensity to the maximum achievable self-susceptibilities, see \underline{Supporting Text}, Supplementary Fig S5.}

At odds with $\ovl{\avg{\chi}}$, however, $\ovl{\chi}_{\max}$ is strongly \red{context-dependent}, being modulated by both BH and (more significantly) TH. This finding agrees with the proposed role of kinetic heterogeneities in creating favourable paths in the miRNA-RNA network through which perturbations can efficiently propagate, as discussed e.g. in  \cite{nitzan,martirosyan}.

\subsection*{\red{Crosstalk becomes more selective upon increasing heterogeneity}}

\red{Along with a higher potential for propagation, increased TH makes crosstalk more selective by systematically involving a smaller number of targets, both in terms of in-coming regulation (i.e. of the number of different transcripts that can regulate a given RNA) and, more significantly, in terms of out-going regulation (i.e. of the number of different transcripts that are regulated by a given RNA). To quantify this aspect, we evaluated the quantities
\begin{gather}
S_{\text{in}} = \frac{1}{N}\sum_{i=1}^N g_i\qquad, \qquad g_i= \frac{\sum_{j\neq i}^{1,N}\chi_{ij}^2}{(\sum_{j\neq i}^{1,N} \chi_{ij})^2}~~;\\
S_{\text{out}} = \frac{1}{N}\sum_{j=1}^N h_j~~,\qquad, \quad h_j= \frac{\sum_{i\neq j}^{1,N} \chi_{ij}^2}{(\sum_{i\neq j}^{1,N} \chi_{ij})^2}~~.
\label{eq:Sin_out}
\end{gather}
Both $g_i$ and $h_j$ vary between 0 and 1, as do $S_{\text{in}}$ and $S_{\text{out}}$. A value $g_i \simeq 0$ indicates that a large number of RNA species can almost equally affect the steady state of RNA species $i$, whereas a value of $g_i\simeq 1$ indicates that RNA $i$ is regulated by a small number of other RNA species. Likewise, when $h_j \simeq 0$ a perturbation of the transcription rate of RNA $j$ affects the steady state of a large number of other RNA species almost equally, while if $h_j \simeq 1$ RNA $j$ only affects a small number of other RNAs. \red{In turn, $g_i^{-1}$ and $h_i^{-1}$ provide an indication of the number of upstream and, respectively, downstream miRNA-mediated contacts of a given RNA species. If follows that} $S_{\text{in}}$ (respectively $S_{\text{out}}$) represents the average of the inverse number of RNAs, a perturbation of which can considerably affect the level of a given RNA (resp. whose level can be affected by a perturbation of a given mRNA).  We will call $S_{\text{in}}$ the {\it incoming  selectivity} and $S_{\text{out}}$ the {\it outgoing selectivity}, respectively. }

\red{Fig \ref{fig3} displays the inverse incoming (Fig \ref{fig3}a) and outgoing (Fig \ref{fig3}b) selectivities as functions of the mean miRNA transcription rate $\overline{\beta}$ for the CLASH interactome. 
\begin{figure*}
\begin{center}
\includegraphics[width = \textwidth]{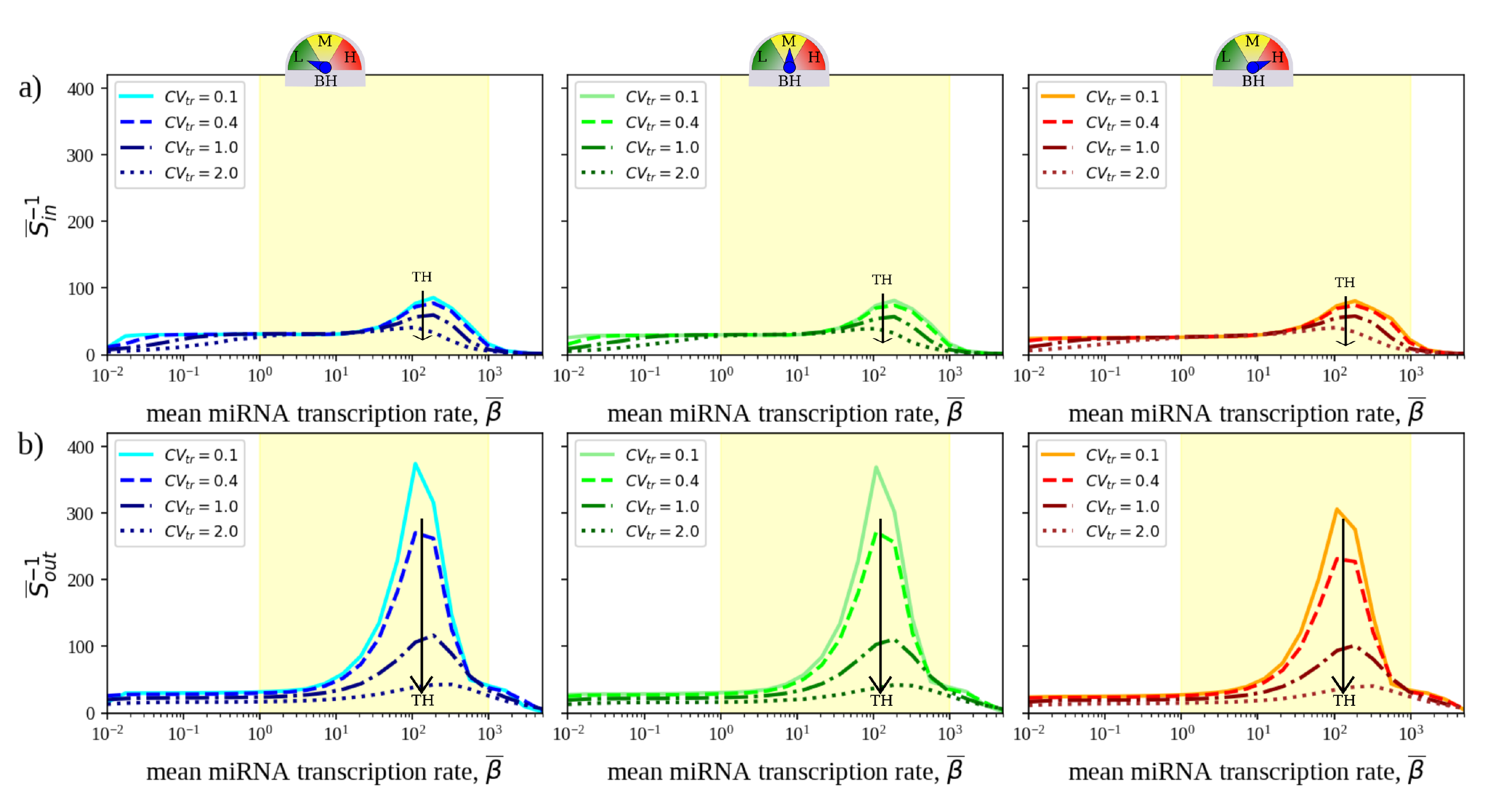}
\end{center}
\caption{\red{\textbf{Crosstalk selectivity in the CLASH interactome.} \textbf{(a)} Inverse of incoming and \textbf{(b)} outgoing selectivities as functions of $\overline{\beta}$ for varying degrees of TH (different curves in the same panel) and BH (reported by the 3-state gauge in different panels). Curves are averaged over 100 independent realizations of transcription rate profiles.}}
\label{fig3}
\end{figure*}
One sees that the typical number of crosstalk partners is modulated significantly only in the susceptible region. Higher degrees of transcriptional heterogeneity in particular tend to make crosstalk increasingly more selective (i.e. to lower $S^{-1}_{\text{in}}$ and $S^{-1}_{\text{out}}$). On the other hand, different degrees of binding heterogeneity (BH) appear to impact this scenario rather weakly.}

\red{Because the selectivity is ultimately a global property tied to susceptibility distributions, a few characteristics of these curves can be understood from features of the latter. For instance, the high selectivity achieved outside the susceptible region is likely due to the existence of strongly crosstalking pairs, enhanced by transcriptional heterogeneities (cf. Fig \ref{fig2}c). Peak inverse selectivity is instead achieved when susceptibility distributions tend to become more homogeneous (cf. Fig S4). Likewise, transcriptional heterogeneity makes distributions less homogeneous, thereby increasing the selectivity. On the other hand, the divergent behaviour of incoming and outgoing components is harder to understand based on these aspects alone, as it possibly involves topological ingredients.}

\red{Keeping in mind that susceptibilities $\chi_{ij}$ are not {\it a priori} symmetric, one can also quantify the degree of asymmetry in terms of the mean relative difference between $\chi_{ij}$ and $\chi_{ji}$. Results (see \underline{Supporting Text}, Section 3) show that, in the CLASH network, a properly defined asymmetry index is robustly maximized in the susceptible regime, where it can achieve a significantly high value that is weakly modulated by TH.}

\red{Summing up, while the behaviour of the mean crosstalk intensity appears to be hard-wired in the topology of the CLASH interactome, other features are tuned by the degree of heterogeneity. Most notably, crosstalk gets stronger and more selective as transcription rates become more diverse, while binding heterogeneities appear to specifically affect the maximum crosstalk intensity achievable. Finally, when crosstalk is strongest, individual crosstalk interactions tend to become more asymmetric, i.e. $\chi_{ij}$ and $\chi_{ji}$ are typically different. As this feature is observed independently of the degree of TH, the emergence of directional crosstalk appears to be an inherent property of miRNA-RNA networks.}

\subsection*{Transcriptional heterogeneities elicit non-local RNA crosstalk}

\begin{figure*}
\begin{center}
\includegraphics[width=0.8\textwidth]{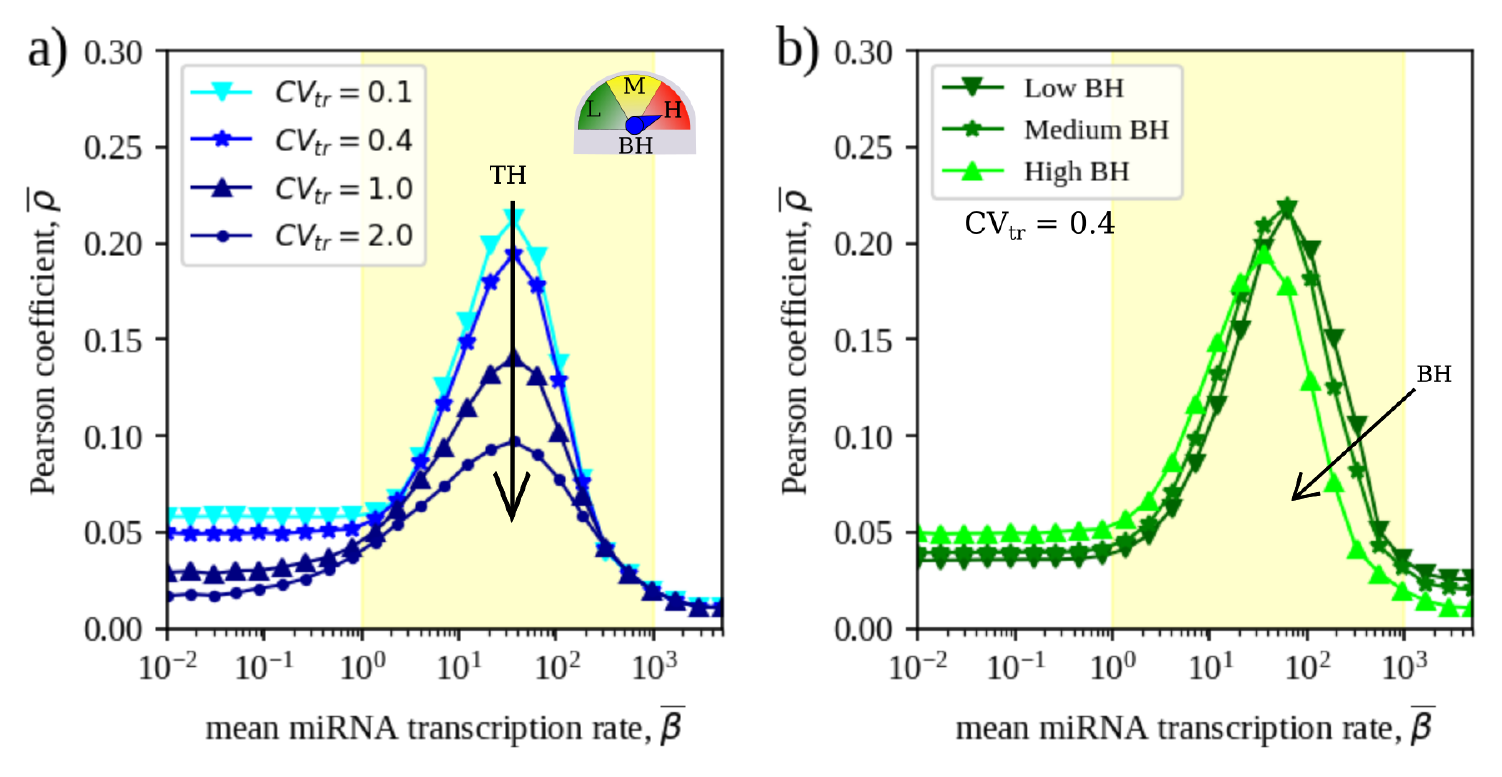}
\end{center}
\caption{\textbf{Non-local effects in the RNA crosstalk scenario derived for the CLASH interactome.} \textbf{(a)} Mean Pearson coefficient $\ovl{\rho}$ quantifying the degree of locality of crosstalk patterns as a function of the mean miRNA transcription rate $\ovl{\beta}$ for different degrees of TH. \textbf{(b)} Behaviour of $\ovl{\rho}$ as a function of $\ovl{\beta}$ in different BH scenarios for a given TH scenario ($\CV_\b=0.4$). Results are obtained in the high-BH scenario by averaging over 100 independent realizations of transcription rates for each degree of TH. In each case, the standard error of the mean is equal to or smaller than the size of the markers.}
\label{fig4}
\end{figure*}

Increased crosstalk intensity and selectivity are accompanied by the establishment of non-local effects, represented by strong effective interactions coupling RNAs that are separated by more than one miRNA species in the miRNA-RNA network. This phenomenon has been addressed e.g. in \cite{nitzan,martirosyan4} in the context of small motifs. To quantify it  in a large-scale network, we consider the quantity
\beq
K_{ij} \equiv \frac{1}{M}\sum_{a=1}^M \frac{1}{\mu_{ia}^0}\frac{1}{\mu_{ja}^0}~~.
\eeq
\red{By definition, $K_{ij}$ is non-zero only if RNAs $i$ and $j$ are co-targeted by at least one miRNA species, while it vanishes for pairs $(i,j)$ that do not share a miRNA regulator.} (About 90\% of potentially crosstalking RNA pairs involves species that are not co-regulated in the CLASH interactome.) In brief, as $\mu^0_{ia}$ is inversely proportional to the binding affinity between miRNA $a$ and RNA $i$, larger values of $K_{ij}$ imply a stronger crosstalk potential between RNA species $i$ and $j$ based only on the network's {\it local} interaction structure and kinetic parameters. If crosstalk mostly occurs between co-regulated RNAs one should therefore expect the pattern of susceptibilities to match that of $K_{ij}$s, at least qualitatively. We hence focus on the Pearson correlation coefficient between the $K_{ij}$s and the susceptibilities $\chi_{ij}$s, i.e. 
\beq\label{rho}
\rho = \frac{\avg{\chi_{ij}K_{ij}} - \avg{\chi_{ij}}\avg{K_{ij}}}{\sqrt{\left(\avg{\chi_{ij}^2} - \avg{\chi_{ij}}^2\right)\left(\avg{K_{ij}^2} - \avg{K_{ij}}^2\right)}}~~~~~~~(0\leq \rho\leq 1)~~.
\eeq
\red{By construction, $\rho\simeq 1$ when RNA crosstalk occurs mainly between co-regulated RNA species, while it gets smaller as the number of non-neighbouring targets that significantly crosstalk increases. Hence $\rho$ effectively quantifies the degree of non-locality in crosstalk patterns (higher $\rho$ implying more local crosstalk).}

Fig~\ref{fig4} shows the behaviour of $\ovl{\rho}$, the average being over realizations of TH. While the correlation peaks in the susceptible region, crosstalk patterns generically appear to  correlate poorly with local topology in the CLASH interactome, as $\ovl{\rho}\lesssim 0.2$. Most notably, $\ovl{\rho}$ decreases significantly as TH is strengthened. \red{This implicates kinetic heterogeneities in the establishment of extended interaction paths that reduce the effective diameter of the interactome by connecting distant RNAs via miRNA-mediated interactions.} In this respect, miRNAs appear to operate on RNAs both as specific repressors of individual transcripts and as a diffuse regulatory layer affecting the transcriptome as a whole.

The most marked effect induced by binding heterogeneities consists in an increase of $\ovl{\rho}$ at small $\ovl{\beta}$ and a shift of the peak correlation at smaller values of $\ovl{\beta}$. Interestingly, changes appear only when the full-fledged variability of binding sites is considered (high BH), while both the homogeneous case (low BH) and the case in which only k-mer and non-k-mer interactions are distinguished (medium BH) return  very similar results. The particular structure of non-k-mer interactions reported in the CLASH data therefore only seems to bear a weak impact on the structure of crosstalk patterns. 

Fig~\ref{fig4} has an important practical implication:  relying on local kinetic parameters like $\mu^0_{ia}$ (or, equivalently, on the binding affinity $k_{ia}^+$) to predict crosstalk interactions could be ineffective due to the significant long-range crosstalk that emerges as the network becomes more and more heterogeneous, especially in terms of transcription rates. This conclusion is most relevant in the susceptible regime, where cells presumably operate and RNA levels are more sensitive to changes in miRNA levels. 

\begin{figure*}[t!]
\begin{center}
\includegraphics[width=0.8\textwidth]{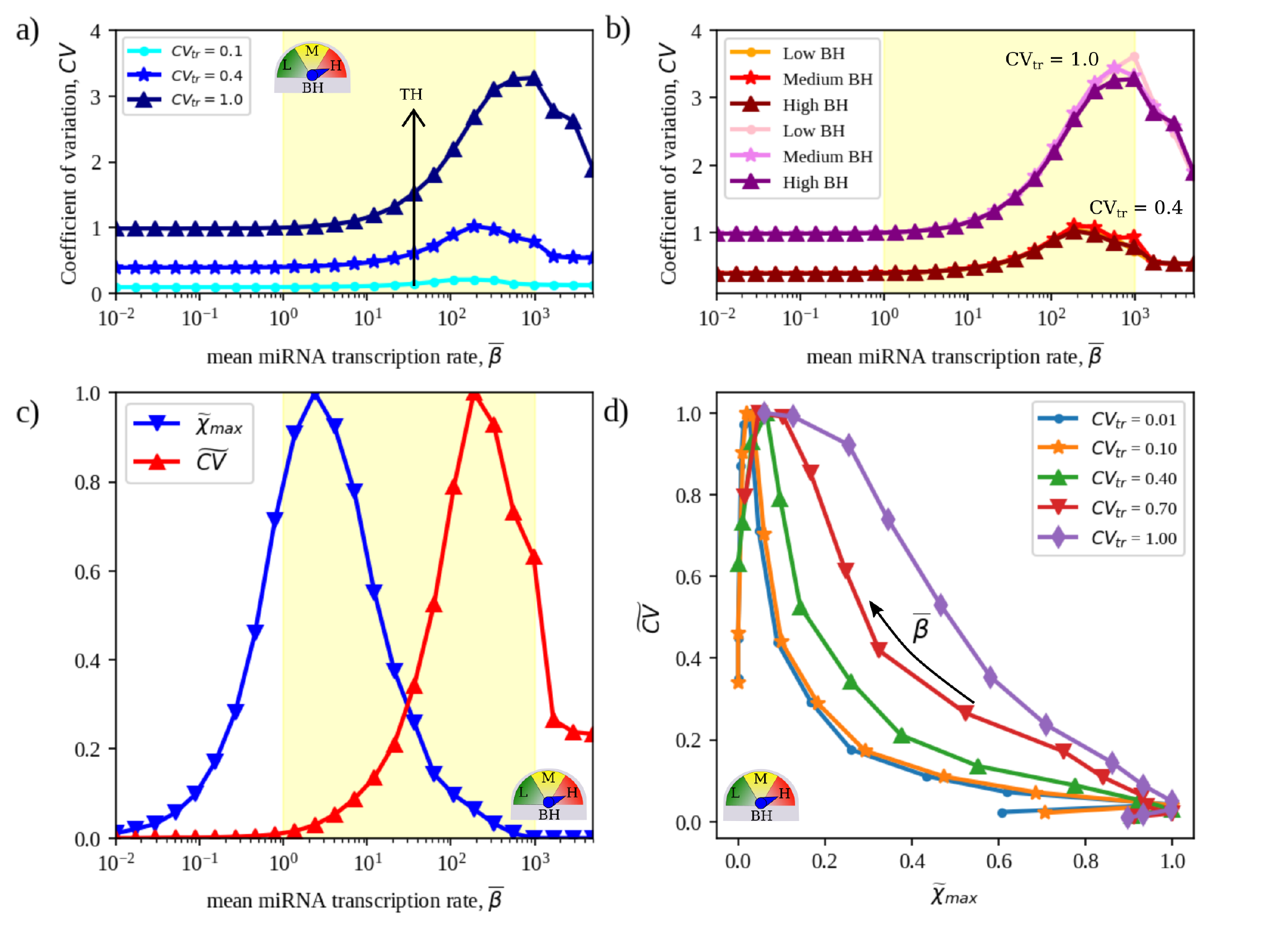}
\end{center}
\caption{\textbf{Robustness of expression profiles from the CLASH interactome in the presence of crosstalk.} \textbf{(a)}  Coefficient of Variation ($CV$) of RNA levels as a function of the overall mean miRNA transcription rate $\ovl{\beta}$ for different degrees of TH. \textbf{(b)} Behaviour of the $CV$ as a function of $\ovl{\beta}$ in different BH scenarios in a fixed TH scenario ($\CV_\b=0.4$). \textbf{(c)} Comparison between the rescaled normalized maximal susceptibility $\widetilde{\chi}_{\max}$ (varying between 0 and 1) and the rescaled normalized Coefficient of Variation  $\widetilde{CV}$ as a function of the overall mean miRNA transcription rate $\ovl{\beta}$ at fixed TH ($\CV_\b=0.4$) and BH (high). \textbf{(d)} $\widetilde{\chi}_{\max}$ vs $\widetilde{CV}$ for different degrees of transcriptional heterogeneity ($\CV_\b$) and high BH. Results are obtained by averaging over 1000 independent TH realizations. In each case the standard error of the mean is equal to or smaller than the size of the markers.}
\label{fig5}
\end{figure*}

\subsection*{More robust expression profiles are associated to stronger RNA crosstalk}

\red{By controlling the availability of their targets, miRNAs  effectively process the variability induced by RNA transcription rates. In some cases (e.g. in presence of specific patterns of correlation between transcription rates), fluctuations can be reduced leading to more finely tuned expression levels \cite{figliuzzi,martirosyan,martirosyan2,re}. In general, though, crosstalk tends to amplify target variability, especially when different species are transcribed independently \cite{figliuzzi}. The exact relationship between crosstalk intensity and transcript variability in extended networks is however bound to depend on the specific features of the crosstalk patterns.}

In Fig~\ref{fig5}a-b we show the coefficient of variation of RNA levels, averages being taken over many independent realizations of TH, as a function of the mean miRNA transcription rate $\ovl{\beta}$ in different BH scenarios. Relative fluctuations exhibit a maximum at large values of $\ovl{\beta}$ within the susceptible region and generically increase with the degree of TH. Variability in transcription rates therefore expectedly promotes variability in the resulting expression profiles. However the increase of fluctuations with respect to the unregulated case ($\ovl{\beta}\to 0$) is negligible or very modest in a broad range of values of $\ovl{\beta}$ within the susceptible region. On the other hand, at fixed TH, different BH scenarios do not appear to affect the robustness of expression profiles (see Fig~\ref{fig5}b). 

Recalling the behaviour of the maximal susceptibility $\ovl{\chi}_{\max}$  (see Fig \ref{fig2}b), one notices that the strongest maximal crosstalk is associated to more robust expression profiles within the susceptible region and, vice-versa, stronger fluctuations in expression profiles occur when crosstalk gets weaker (see Fig \ref{fig5}c). In other terms, uncorrelated transcriptional heterogeneities tend to be amplified when crosstalk is suppressed (higher miRNA expression levels), while they are more efficiently \red{contained} when the strongest crosstalk emerges. This scenario is summarized in Fig \ref{fig5}d: for any given degree of TH, as miRNA availability increases, crosstalk intensity on one hand and fluctuations of the output levels on the other are subject to a tradeoff that gets stronger as transcription rates becomes more homogeneous. 

These results clearly implicate transcriptional heterogeneities as a key determinant of the stability of expression profiles even in presence of crosstalk, in line with previous observations on small networks \cite{re,martirosyan}. It is however important to remark that this picture was obtained under the assumption of uncorrelated extrinsic fluctuations in RNA transcription rates. The presence of correlations might considerably alter this conclusion, as was first discussed in \cite{figliuzzi}.

\begin{figure*}
\begin{center}
\includegraphics[width=\textwidth]{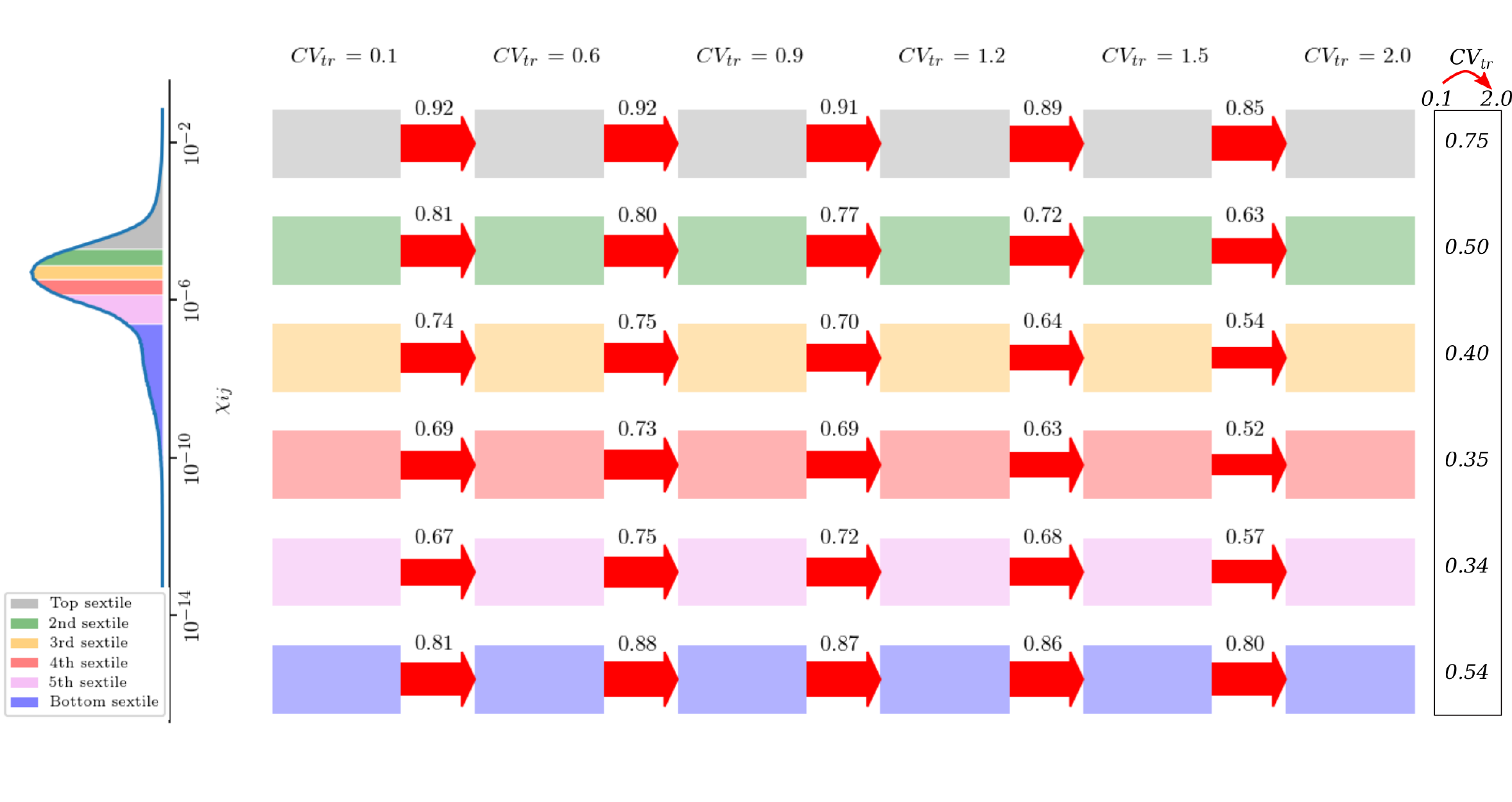}
\end{center}
\caption{\textbf{Global robustness of crosstalk patterns from the CLASH interactome with increasing degrees of transcriptional heterogeneity.} For each susceptibility sextile, the marking above the arrow reports the fraction of crosstalk interactions that are preserved upon increasing the degree of TH. \red{The last column reports the fraction of interactions that are conserved passing from the lowest to the highest degree of TH.} Results are obtained by averaging over 100 independent TH realizations, assuming high BH and mean miRNA transcription rate $\ovl{\beta}=30$ (deep in the susceptible regime). Different intermediate values of $\ovl{\beta}$ return qualitatively identical scenarios.}
\label{fig6}
\end{figure*}

\subsection*{Crosstalk patterns are resilient to transcriptional heterogeneity}


\red{After analysing systemic properties, we now ask to what degree crosstalk patterns are preserved upon increasing the level of transcriptional heterogeneity. A global analysis shows (see Fig \ref{fig6}) that susceptibilities are remarkably well preserved statistically as the degree of transcriptional heterogeneity increases. Most notably, about 75\% of the RNA pairs that are in the top sextile for crosstalk intensity at the lowest $\CV_\b$ ($\CV_\b=0.1$) persist in the top sextile when TH is 20-fold larger ($\CV_\b=2$). Such a fraction is larger than would be expected by chance (about 58\%), implying the existence of a significant backbone of RNA-RNA interactions resilient to transcriptional heterogeneity. A similar picture holds for the other sextiles. It is also instructive to inspect robustness specifically for RNA pairs that do not share any miRNA regulators, which amount to roughly 90\% of the total. Fig S6 in \underline{Supporting Text} shows that, even for such `distant' RNAs, about 73\% of the most strongly interacting pairs are conserved across all degrees of TH. For reference, the 15 most strongly interacting pairs overall and among distant RNAs are displayed in \underline{Supporting Text}, Figs S7 and S8. (Notice that distant pairs carry a susceptibility that is two orders of magnitude smaller than the maximum but two orders of magnitude larger than the average.)}

\red{Weak sensitivity to changes in transcriptional heterogeneity would be expected if crosstalk interactions were functionally significant. Remarkably, this appears to be the case across a broad range of degrees of TH, both for short-range (mediated by a single miRNA species) and long-range (resulting from extended  miRNA-mediated chains) crosstalk interactions.}

\begin{figure*}
\begin{center}
\includegraphics[width=\textwidth]{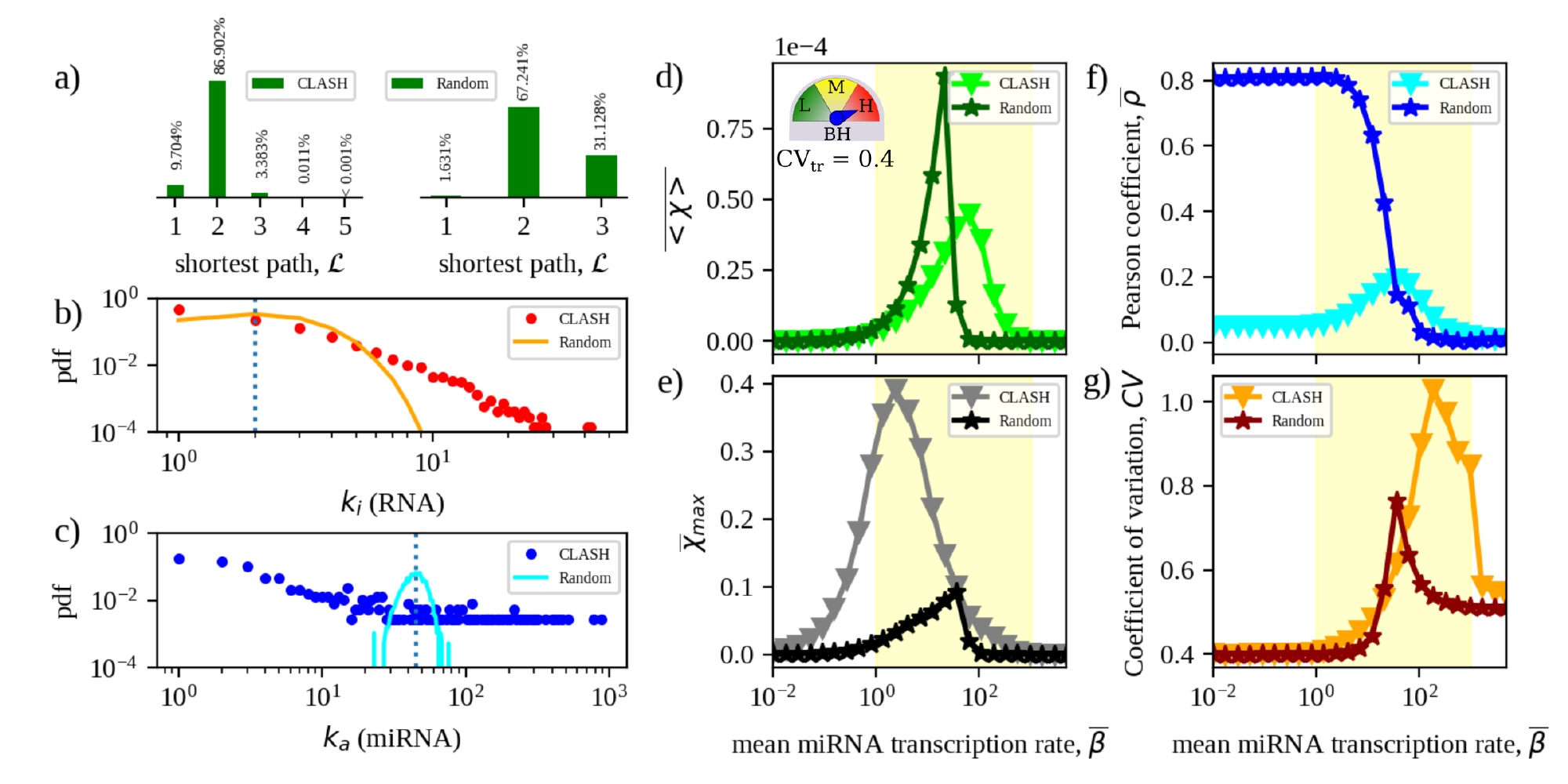}
\end{center}
\caption{\textbf{Comparison between RNA crosstalk in the CLASH interactome and its randomized counterparts.} \textbf{(a)} 
Frequency of shortest miRNA-mediated paths between RNA species. The length corresponds to the minimum number of miRNA species that can mediate a crosstalk interaction between two RNA species. \textbf{(b, c)} Distributions of node degrees for RNAs (top) and miRNAs (bottom). \textbf{(d, e)}  Mean and maximum susceptibilities as a function of $\ovl{\beta}$. \textbf{(f, g)} Pearson coefficient $\ovl{\rho}$ quantifying crosstalk locality and coefficient of variation of the levels of free RNAs as functions of $\ovl{\beta}$. In panels d through g, TH is fixed at $\CV_\b= 0.4$ and averages performed over 1000 independent realizations of TH under high BH. In each case, the standard error of the mean is equal to or smaller than the size of the markers.}
\label{fig7}
\end{figure*}

\subsection*{Node degrees are the key topological determinants of the crosstalk scenario in the CLASH interactome}
\label{sec:randNet}

To appraise the role of the specific wiring encoded by the CLASH data in determining the scenario described so far, we compared our results against a null model obtained by randomly re-wiring the CLASH interactome. Specifically, we re-assigned each link to a randomly chosen miRNA-RNA pair, thereby preserving only the overall numbers of links and nodes while altering all other topological features like node degrees, degree-degree correlations, etc (see Methods). Each independent re-wiring process leads to a different final network. \red{These randomized versions diverge from the original miRNA-RNA network in two basic aspects.} In first place, they are slightly more compact, as evidenced by the distribution of the shortest miRNA-mediated paths between any two RNAs shown in Fig \ref{fig7}a. In addition, the randomization alters the distribution of node degrees by effectively eliminating the most highly connected RNA and miRNA species that are found in the CLASH data (see Fig \ref{fig7}b and \ref{fig7}c). Results obtained for key crosstalk descriptors in the CLASH and randomized networks (averaged over many realizations of the randomization protocol) are illustrated in Fig \ref{fig7}d--g.

Randomized networks display a much larger (about two-fold) mean susceptibility for crosstalk than the CLASH interactome, possibly due to the fact that miRNA targets are generically closer in the randomized versions. However, the maximum achievable crosstalk strength $\ovl{\chi}_{\max}$ is about 4 times smaller in the random networks compared to CLASH. Moreover, the susceptibility profile is more concentrated in the randomized network than it is for the CLASH network, reflecting a significantly narrower susceptible region (see \underline{Supporting Text}, Fig S9). \red{Naturally selected miRNA-RNA networks therefore appear to foster the emergence of stronger crosstalk links.} In addition, the Pearson coefficient $\ovl{\rho}$ quantifying the linear correlation between susceptibilities and local interaction parameters attains a much larger value in the randomized network with respect to the CLASH interactome throughout most of the susceptible region (see Fig \ref{fig6}f). miRNA-mediated crosstalk in random networks is therefore significantly more local, and thereby easily predictable e.g. by miRNA-RNA affinities, than it is in a network shaped by natural selection. Finally, expression profiles generated in the randomized network are slightly more stable than those found in the CLASH interactome (as quantified by the coefficient of variation, see Fig \ref{fig6}f). This feature is however more marked at higher miRNA expression levels, where RNA crosstalk is generically weaker. The basic traits of the RNA crosstalk emerging in randomized versions of the CLASH data are hence substantially different from those characterizing the interactome. \red{\underline{Supporting Text}, Section 3 and Fig S10 report the behaviour of the asymmetry and selectivity indices in randomized networks. At odds with the results obtained for the interactome (for which the asymmetry is weakly dependent on parameter heterogeneity), crosstalk in randomized networks becomes drastically more asymmetric and selective with increasing degrees of TH, although the number of interaction partners is generically higher in the randomized topology than it is in the interactome. In other terms, such features appear to be less robust to parameter heterogeneity in random structures than they are in naturally selected networks.}

\red{Note that, by applying a more conservative protocol that reshuffles miRNA-RNA links while preserving node degrees, one retrieves a crosstalk scenario that is essentially identical to that found for the original CLASH interactome (see \underline{Supporting Text}, Section 4). This indicates that degree sequences (i.e. the topology of miRNA-RNA interactions encoded by the different types of couplings), as opposed to e.g. degree-degree correlations or other higher-order topological features, are the key geometric controllers of RNA crosstalk patterns. Enhanced crosstalk and non-locality therefore appear to be encoded by selection within the structure of the miRNA-RNA network interaction.}

\begin{figure*}
\begin{center}
\includegraphics[width=0.8\textwidth]{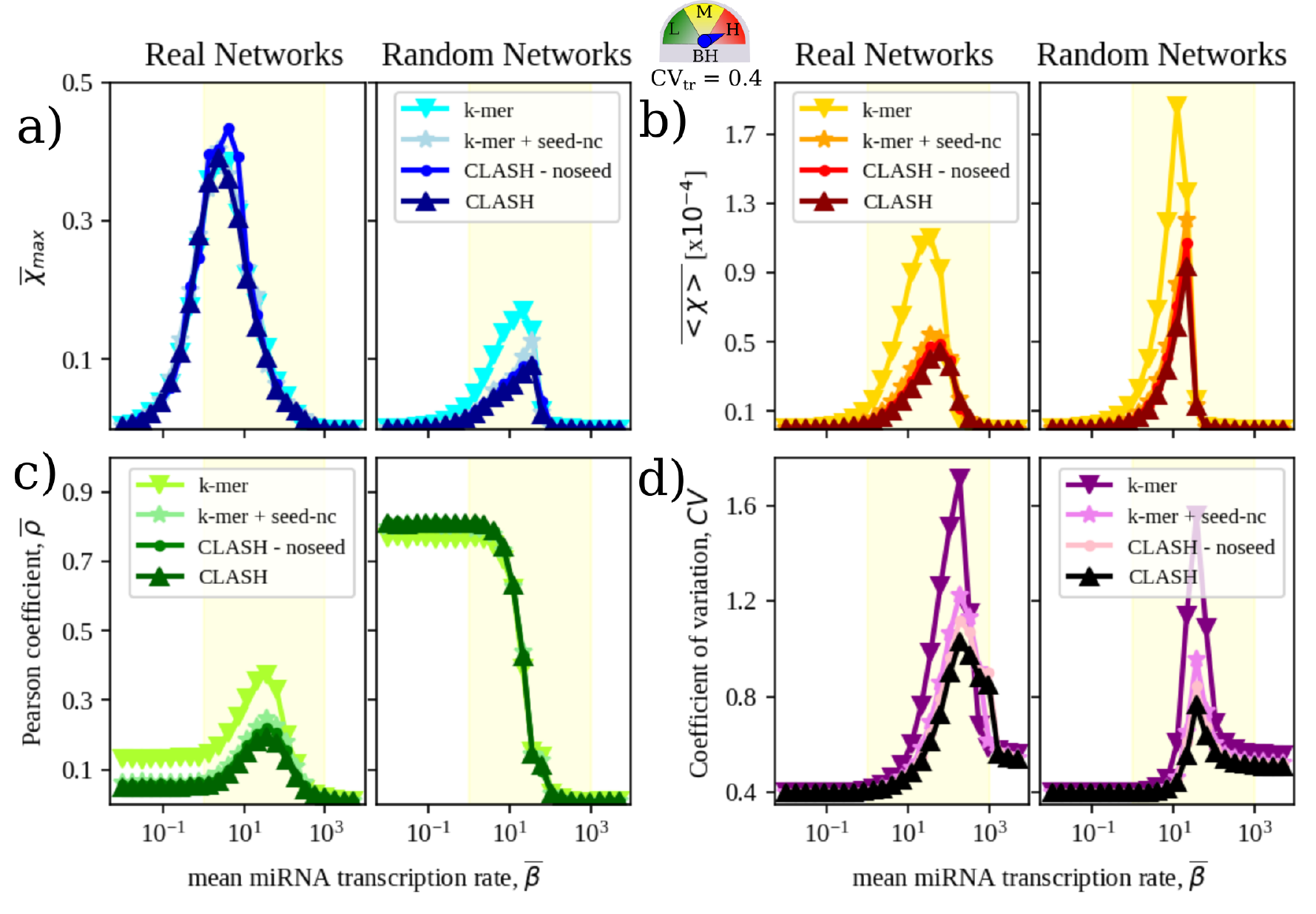}
\end{center}
\caption{\textbf{Crosstalk features in k-mer-based CLASH subnetworks and in their randomized counterparts.} \textbf{(a, b)} Mean and maximum susceptibilities as functions of $\ovl{\beta}$. \textbf{(c)} Pearson coefficient $\ovl{\rho}$ quantifying crosstalk locality. \textbf{(d)} Coefficient of Variation of the levels of free RNAs as a function of $\ovl{\beta}$. The degree of TH is fixed at $\CV_\b= 0.4$ (with averages performed over 1000 independent realizations of TH) and high BH is assumed. In each case, the standard error of the mean is equal to or smaller than the size of the markers. The term `CLASH-noseed' indicates the full CLASH interactome except for `noseed' type of interactions.}
\label{fig8}
\end{figure*}

\subsection*{Canonical and non-canonical binding sites control different aspects of RNA crosstalk}

\red{A key question at this point is whether the observed crosstalk scenario is mainly due to the canonical (stronger) k-mer pairings or, rather, if non-canonical (weaker) binding sites contribute to its establishment. A breakdown of the topology of the subnetworks induced by the different classes of interactions in the CLASH data shows significant similarities (see \underline{Supporting Text}, Fig S11). Based on topology alone, then, appraising the role of non-canonical interactions is not simple.}

To clarify this point, one can repeat the above analysis by successively adding each type of pairings shown in Fig \ref{fig1}c to the subnetwork induced by k-mer interactions in the CLASH data. After evaluating susceptibilities in each case, one sees (see Fig \ref{fig8}) that the crosstalk scenario underlied by the k-mer layer is qualitatively similar to that retrieved for the complete CLASH interactome. In particular, the k-mer network alone expectedly suffices to explain the maximum achievable crosstalk with quantitative accuracy. However, k-mer interactions by themselves would yield stronger mean susceptibility, slightly more local crosstalk patterns and significantly larger variability of output profiles compared to the full network. Perhaps surprisingly, each of these aspects therefore appear to be quantitatively modulated to some degree by the weaker non-canonical interactions. 

To further validate this picture, we have analyzed the RNA crosstalk scenario in the cancer-specific interactomes reconstructed in \cite{chiu}. \red{These networks comprise canonical pairings only and their} basic topological characteristics are noticeably distinct from those found in CLASH (see \underline{Supporting Text}, Fig S12). Results are summarized in \underline{Supporting Text}, Fig S13. The low-$\ovl{\beta}$ behaviour starkly contrasts with that found in the CLASH reconstruction, in that crosstalk carries a stronger local component. In addition, maximal susceptibilities are roughly four times weaker in these networks. As a consequence, crosstalk is generically attenuated compared to CLASH and the potential to process (amplify) transcriptional heterogeneities is limited. Output profiles are consequently more stable against transcriptional variability across the whole range of levels of miRNA expression. These differences aside, the emergent crosstalk pattern robustly shows enhanced maximal intensity and non-locality with respect to their randomized counterpart, in qualitative agreement with the emergent crosstalk picture derived for the CLASH interactome.

\section*{Discussion}

\subsection*{Methodological choices}

The system-level crosstalk scenarios studied here were derived under a few key methodological choices. First, we focused on the steady states of the mass action kinetics of miRNA-RNA interactions, Eq (\ref{ssm}). \red{While reasonable for timescales of the order of $1/d$ and $1/\delta$ (and compatible with those analyzed in the experimental literature), this choice} excludes from our analysis the rich phenomenology observed during transients \cite{figliuzzi2}, when crosstalk can be amplified over timescales determined by the details of the interaction kinetics. \red{Likewise, we can't account directly for intrinsically dynamic regulatory mechanism like the dependence of miRNA decay rates on the round of recycling observed in \cite{baccarini}.} Second, we opted to set a few parameters to values compatible with experimental evidence (see Table \ref{tab:data}) while treating miRNA and RNA transcription rates as independent, identically distributed quenched random variables with prescribed probability distributions. All our results were then obtained by averaging over many realizations of such parameters. Such an approach highlights features of the emerging crosstalk patterns that may be expected to be independent of the specific choice of transcription rates. \red{On the flip side, we are unable to characterize crosstalk for specific, possibly more realistic choices of transcription rates.} Finally, we used the mean miRNA transcription rate as the only control parameter. While we explored a very broad range of values thereof, the physiologically relevant regime is likely to lie at intermediate miRNA transcription intensities, where RNA levels are more sensitive to changes in miRNA levels. 

By assuming constant mean transcription rates we are effectively discarding the possibility that different RNAs or miRNAs are transcribed at very different rates \red{(at least at low enough values of $\CV_\b$). Significant inhomogeneities in the average biosynthesis rates of individual molecular species may affect our results. Yet, the highly interconnected structure of miRNA-RNA networks and the hierarchical organization of miRNA-RNA binding affinities \cite{rzepiela} suggest that RNA crosstalk could be more influenced by global miRNA availability than by the specific structure of the miRNA population, at least in standard physiological conditions (e.e. in absence of strong miRNA induction).} This is ultimately the scenario we probe in our study. \red{Likewise, we are ignoring the possibility that heterogeneity parameters are correlated. As mentioned before, correlations between transcription rates would alter the picture regarding the processing of fluctuations \cite{figliuzzi}. Likewise, correlations between topological and transcriptional parameters, like those observed in \cite{narang}, would  the effects of heterogeneity, thereby significantly affecting crosstalk patterns.}

To test the degree to which selection-shaped features of the miRNA interactome influence the emergent crosstalk pattern, we then studied how effective interactions are modulated by the structure of miRNA-RNA binding strengths and by the specific wiring encoded in data. In the former case, we were interested in evaluating the  relevance for crosstalk of non-canonical binding sites, whose repressive efficiency is likely very limited \cite{agarwal}. In the latter, we aimed instead at understanding (i) whether crosstalk in naturally selected miRNA-RNA networks is qualitatively different from that arising in random networks and, if so, (ii) which topological features of real networks underlie the observed disparities.

\subsection*{Outlook}

In broad terms, our analysis shows that RNA crosstalk in an extended network is modulated by miRNA availability both in terms of its \red{basal level measured by the mean intensity} and in terms of its maximal achievable strength. The typical crosstalk interaction is relatively weak. In specific, it is roughly four orders of magnitude smaller than the mean self-susceptibility, i.e. the mean change in the level of free transcript induced by a variation of its own transcription rate. Still, a multitude of strong crosstalk interactions arise, whose intensity is comparable to that of self-interactions. This in turn generates highly selective and directional crosstalk patterns. Notably, while co-regulated RNAs typically bear the strongest crosstalk links, non-co-regulated, or distant, RNAs can still crosstalk at significant intensities (roughly two orders of magnitude above the \red{basal level}). In such conditions, the typical RNA-to-RNA distance, in terms of number of links of the miRNA-mRNA interaction networks, above which one species can be considered to be insensitive to perturbations carried out on another species becomes comparable to the diameter of the network. A perturbation in the transcription level of one RNA can then be broadcast (via a chain of miRNA-mediated effective interactions) to distant RNA nodes, potentially propagating over the entire network. \red{Such a feature is intrinsically due to competition and renders local kinetic parameters ineffective priors to predict crosstalk interactions. miRNAs therefore appear to manage a system-level regulatory layer where they operate collectively to mediate a complex, heterogeneous and robust network of RNA cross-regulation. More work is however required to fully unravel its functional capabilities, especially concerning the buffering of fluctuations and gene expression noise.}

\red{The scenario we describe is qualitatively preserved if crosstalk is assumed to be carried by canonical interactions alone. In particular, the latter are highly effective modulators of crosstalk intensity. Non-canonical binding sites however, while substantially ineffective repression-wise, can enhance non-locality thereby extending the crosstalk range. Topological features of the naturally selected interactome were also found to bear a significant effect on crosstalk. Specifically, they lead to a broader susceptible region, higher maximal susceptibility, and more pronounced non-local effects than their randomized counterparts. In this respect, selection appears to have favoured the emergence of such features at system level.}

\red{It is important to stress that the crosstalk interactions on which we focus are quantified by susceptibilities, Eq (\ref{eq:Chi}). This implies that (i) they are driven by competition effects exclusively, and (ii) they are generated by small perturbations of RNA levels (as opposed e.g. to the models of  \cite{rzepiela,chiu}).} Our scenario might therefore be close to a standard, homeostatic physiology in which transcription rates only undergo small variations. In this respect, the emergence of significant non-local effects is a surprising consequence of networking. \red{Large perturbations, like the strong induction of a particular miRNA species, should be expected to amplify the crosstalk picture discussed here. However, responses to perturbations may become non-linear when the applied stimulus exceeds a threshold \cite{figliuzzi2}. In such conditions, susceptibilities or standard correlation coefficients are likely inappropriate to describe crosstalk. More theoretical work on miRNA-RNA networks is required to fully sort out this case.}

\red{Unfortunately, probing the crosstalk scenario we describe in experiments could be challenging essentially due to weakness and non-locality. To validate the picture we describe, both in terms of individual interactions and of global features, one may however resort to transcriptomic data. Recent work has indeed identified a specific group of correlation functions that, under certain conditions, yield excellent approximations for the real susceptibilities \cite{martirosyan4}. Evaluating such quantities on RNA readouts would then provide a direct, data-driven snapshot of RNA crosstalk.}

\subsection*{Conclusion}

\red{Besides their important role as negative controllers of gene expressiom, miRNAs mediate the establishment of extended networks of RNA cross-regulation. Several features of these networks appear to be hard-wired in the topology of the underlying miRNA-RNA interactome, while others are modulated by transcriptional and/or binding heterogeneities. Whereas the typical crosstalk interaction generated by small changes in RNA availability is weak, non-local effects are significant. Crosstalk-based regulation therefore appears essentially as a system-wide phenomenon, enhanced by variability in kinetic parameters. In physiological conditions, such a regulatory layer can potentially contribute to a variety of functions, such as the processing of transcriptional heterogeneities and the coordination of large-scale rearrangements of RNA levels, similar to the responses observed in \cite{rzepiela}. The broader picture we have derived might however apply more generally to networks of molecular species competing for a common resource.}

\section*{Materials and methods}

\subsection*{miRNA interactomes}

For the CLASH interactome, after parsing the original bipartite network derived in \cite{helwak} to remove degeneracies and disjoint nodes, we found $N=6,943$ RNA species (implying about $4.8\times 10^7$ potential crosstalk interactions) and $M=383$ miRNA species connected by 17,411 edges carrying different binding strengths. The same pipeline was applied to the tumor-type specific miRNA-RNA networks obtained in \cite{chiu} and based on the Cupid protocol for predicting microRNA-target interactions \cite{chiu2}, which accounts for canonical pairings exclusively. The resulting miRNA-RNA networks are considerably larger than the CLASH interactome, as evidenced by the comparison of degree distributions given in \underline{Supporting Text}, Fig S12.

\subsection*{Computational analysis}

With parameters set as described, Eq (\ref{chimatrix}) was solved numerically for each of the networks cosidered using Python scripts based on NumPy \cite{vanderwalt} and SciPy \cite{scipy}. \red{The code is available from https://github.com/matmi8/RNAnet.} In presence of TH, results were averaged over multiple independent realizations of the vectors $\mathbf{b}=\{b_i\}_{i=1}^N$ and $\boldsymbol{\beta}=\{\beta_a\}_{i=1}^M$ of RNA and miRNA transcription rates (respectively) for each value of $\CV_\b$. \red{The number of realizations was chosen in each case to ensure a stable estimation of different quantities. Details are given in figure captions. All other parameters, both kinetic and topologic, were kept fixed.} Likewise, in the case of topological heterogeneity, results were averaged over 100 networks obtained by independently randomizing the original miRNA-RNA network while keeping all other parameters, both transcriptional and kinetic, fixed. \red{100 independent randomizations of the interactome sufficed to ensure stable averages in each condition.}

\section*{Acknowledgments}

We are indebted with Matteo Figliuzzi for contributing to the early stages of this work, and gratefully acknowledge Carla Bosia, Marco Del Giudice, Salil Garg and Andrea Pagnani for discussions and suggestions. Work was supported by the European Union's Horizon 2020 Research and Innovation Staff Exchange program MSCA-RISE-2016 under Grant Agreement Nr 734439 (INFERNET).

\clearpage

\renewcommand{\thefigure}{S\arabic{figure}}
\setcounter{figure}{0}    
\setcounter{table}{0}

\section*{Supporting Text}

\subsubsection{Derivation of RNA susceptibilities in generic miRNA-RNA networks}

To derive Eq (6) of the main text, we start recasting the expressions for  $[m_i]$ and $[\mu_a]$ (see Eq.~(2) of main text) as 
\begin{gather}
\label{eq:sup:steady_mn}
\displaystyle [m_i] = \frac{m_i^{\star}}{1 +\displaystyle\sum_{a=1}^{M} \frac{[\mu_a]}{\mu_{ia}^0}} \equiv m_i^{\star} F_i~~,\\
\displaystyle [\mu_a]= \frac{\mu_a^{\star}}{1 + \displaystyle\sum_{i=1}^{N} \frac{[m_i]}{m^0_{ia}}} \equiv \mu_a^{\star} F_a~~.
\end{gather}
Using these, we immediately obtain
\begin{equation}
\label{eq:sup:Xij}
\chi_{ij} \equiv d_i \frac{\partial [m_i]}{\partial b_j} = \frac{[m_i]}{m_i^{\star}} \delta_{ij} + \frac{[m_i]^2}{m_i^{\star}}\sum_{a \in i} \frac{[\mu_a]^2}{\mu^0_{ia}\mu_a^{\star}}\sum_{\ell\in a} \frac{\chi_{\ell j}}{m^0_{\ell a}}~~,
\end{equation}
where we used the identities
\begin{gather}
\label{eq:sup:identities}
\frac{\partial Fi}{\partial[\mu_a]} = -\frac{F_i^2}{\mu^0_{ia}}~~,\\
\frac{\partial F_a}{\partial[m_i]} = -\frac{F_a^2}{m^0_{ia}}~~.
\end{gather}
Eq (\ref{eq:sup:Xij}) can be re-cast in the compact form
\begin{subequations}
\label{eq:sup:matricial1}
\begin{eqnarray}
\sum_{l=1}^N \left( \delta_{il} - W_{il}\right)\chi_{lj} = \frac{m_i}{m_i^{\star}}\delta_{ij}~~,\\
\left( \mathbf{\widehat{1}- \widehat{W}} \right) \boldsymbol{\widehat{\chi}} =\mbox{diag} \left( \mathbf{\frac{m}{m^\star}}\right)~~.
\end{eqnarray}
\end{subequations}
where $\boldsymbol{\widehat{\chi}}$ is the susceptibility matrix (with elements $\chi_{ij}$), $\mbox{diag} \left( \mathbf{\frac{m}{m^\star}}\right)$ denotes the diagonal matrix with elements $\{m_i/m_i^{\star}\}$ while $\mathbf{\widehat{W}}$ is an $N\times N$ matrix with elements 
\beq
\label{eq:sup:W}
W_{ij}\equiv\left(\mathbf{\widehat{W}}\right)_{ij} = \frac{[m_i]^2}{m_i^{\star}}\sum_{a \in (i\cap j)}\frac{1}{m^0_{ja}\mu^0_{ia}}\frac{[\mu_a]^2}{\mu_a^{\star}} ~~.
\eeq
It follows that
\begin{equation}\label{eq:sup:Xij}
\boldsymbol{\widehat{\chi}} = \left( \mathbf{\widehat{1}- \widehat{W}} \right)^{-1} \mbox{diag} \left( \mathbf{\frac{m}{m^\star}}\right) ~~.
\end{equation}
Recalling that, if all eigenvalues of $\mathbf{\widehat{W}}$ are strictly smaller than 1 in absolute values (as is easily verified numerically to the case in this study), one has $\left( \mathbf{  \widehat{1} - \widehat{Z} }\right)^{-1} = \sum_{n\geq 0}	\mathbf{\widehat{Z}}^n $, one finds that
\beq
\label{eq:sup:Xij_exp}
\chi_{ij} = \sum_{n\geq 0} \left(\mathbf{\widehat{W}}^{n} \right)_{ij}  \frac{[m_j]}{m_j^{\star}}  \equiv \sum_{n\geq 0} \chi_{ij}^{(n)}~~.
\eeq 
Expressions \eqref{eq:sup:Xij} and \eqref{eq:sup:Xij_exp} clarify an important point. While $W_{ij}$ is different from zero only if RNAs $i$ and $j$ are co-regulated by at least one miRNA species, the elements of $\mathbf{\widehat{W}}^{n}$ are different from zero if there is at least one chain of $n$ miRNAs joining RNAs $i$ and $j$. In practice, this is what allows for crosstalk to occur even between RNAs that are not directly co-regulated, as shown explicitly within a toy model in the following section.

\subsubsection{Susceptibility between distant RNA pairs}

\begin{figure}[t]
\begin{center}
\includegraphics[width = 0.45\textwidth]{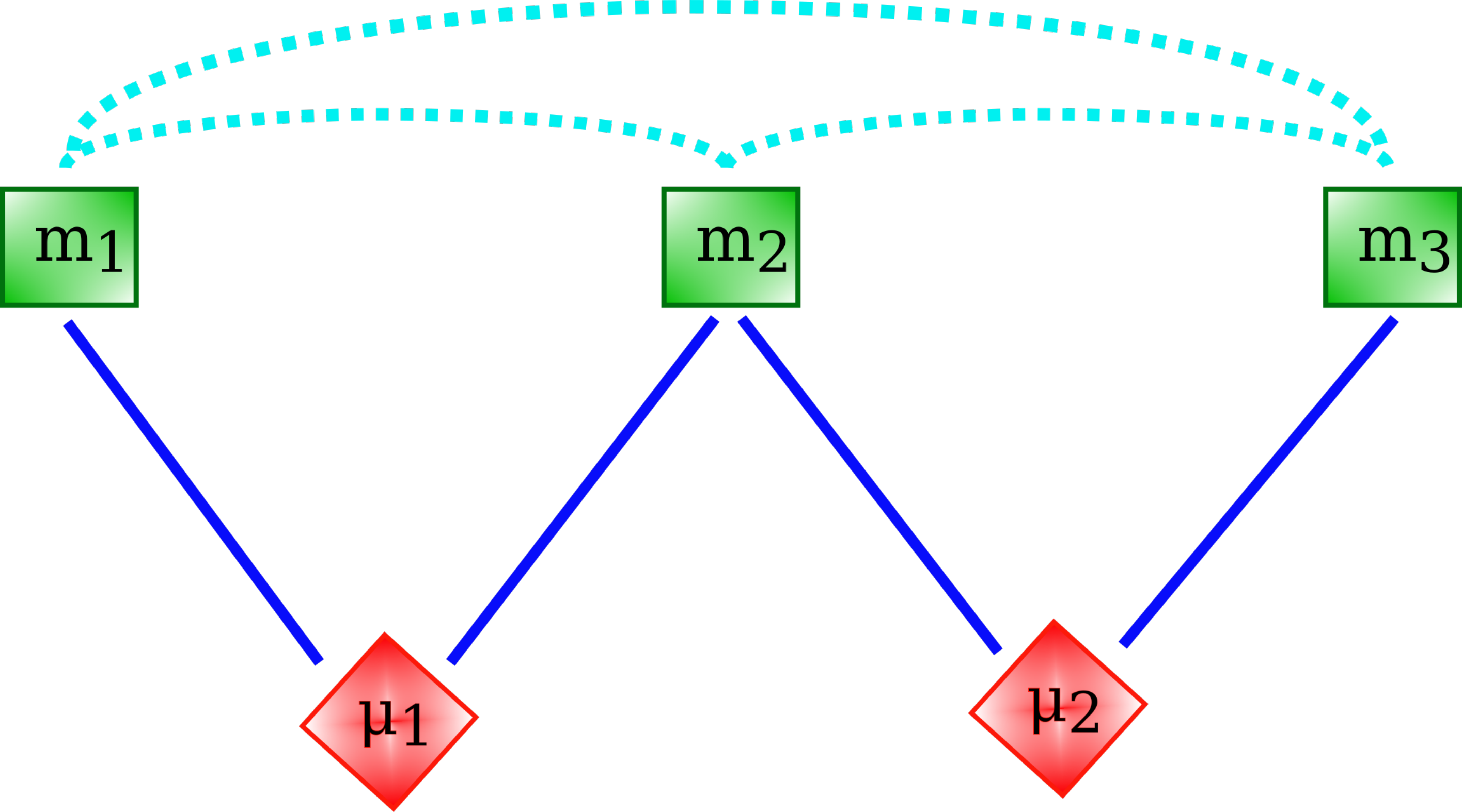}
\end{center}
\caption{\textbf{Toy miRNA-RNA network.} Two miRNA species and three RNA species interact by direct couplings represented by the continuous blue lines. Dotted lines denote instead the effective crosstalk interactions that are established between RNAs as a consequence of competition to bind miRNAs. These correspond in turn to the non-zero elements of the susceptibility matrix $\boldsymbol{\widehat{\chi}}$.}
\label{supp:toy}
\end{figure}

To show explicitly how non-zero susceptibilities can arise between  pairs of RNAs connected by chains of miRNA-mediated couplings from Eq \ref{eq:sup:Xij}, we compute here the susceptibility matrix for a toy network formed by three RNA and two miRNA species,  Fig \ref{supp:toy}. The $\mathbf{\widehat{W}}$ matrix for this network reads
\begin{equation}
\boldsymbol{\widehat{W}} = 
\begin{pmatrix}
w_{11} & w_{12} & 0\\
w_{21} & w_{22} & w_{23}\\
0 & w_{32} & w_{33}\\
\end{pmatrix}~~.
\end{equation}
Two elements ($w_{13}$ and $w_{31}$) are nil since RNAs $1$ and $3$ are not co-targeted by any miRNA. Nevertheless, using (\ref{eq:sup:Xij}), the susceptibility matrix turns out to be given by   
\begin{widetext}
\begin{equation}
\boldsymbol{\widehat{\chi}} = \frac{1}{\text{det} \left( \mathbf{\widehat{1}- \widehat{W}} \right)}
\begin{pmatrix}
(\widetilde{w}_{{22}}\widetilde{w}_{{33}} - w_{23}w_{32} ) \frac{m_1}{m_1^\star}   & ( \widetilde{w}_{{33}}w_{21}) \frac{m_2}{m_2^\star} & (w_{21}w_{32}) \frac{m_3}{m_3^\star}\\
( \widetilde{w}_{{33}}w_{12}) \frac{m_1}{m_1^\star} & (\widetilde{w}_{{11}}\widetilde{w}_{{33}} ) \frac{m_2}{m_2^\star}& ( \widetilde{w}_{{11}}w_{32}) \frac{m_3}{m_3^\star}\\
(w_{12}w_{23}) \frac{m_1}{m_1^\star} & ( \widetilde{w}_{{11}}w_{23}) \frac{m_1}
{m_2^\star}& (\widetilde{w}_{{11}}\widetilde{w}_{{22}} - w_{12}w_{21})\frac{m_3}{m_3^\star}\\
\end{pmatrix}~~,
\end{equation}
\end{widetext}
where $\widetilde w_{ij} = w_{ij} -1 $. Hence a non-zero susceptibility binds RNA species 1 and 3, which are connected by the chain of interactions passing through RNA 2. This connection is also evidenced by the form of the corresponding elements of $\boldsymbol{\widehat{\chi}}$. The above equation also shows explicitly that, in general, $\chi_{ij}$ and $\chi_{ji}$ are different.

\subsubsection{Crosstalk asymmetry}

\begin{figure*}
\begin{center}
\includegraphics[width = 0.85\textwidth]{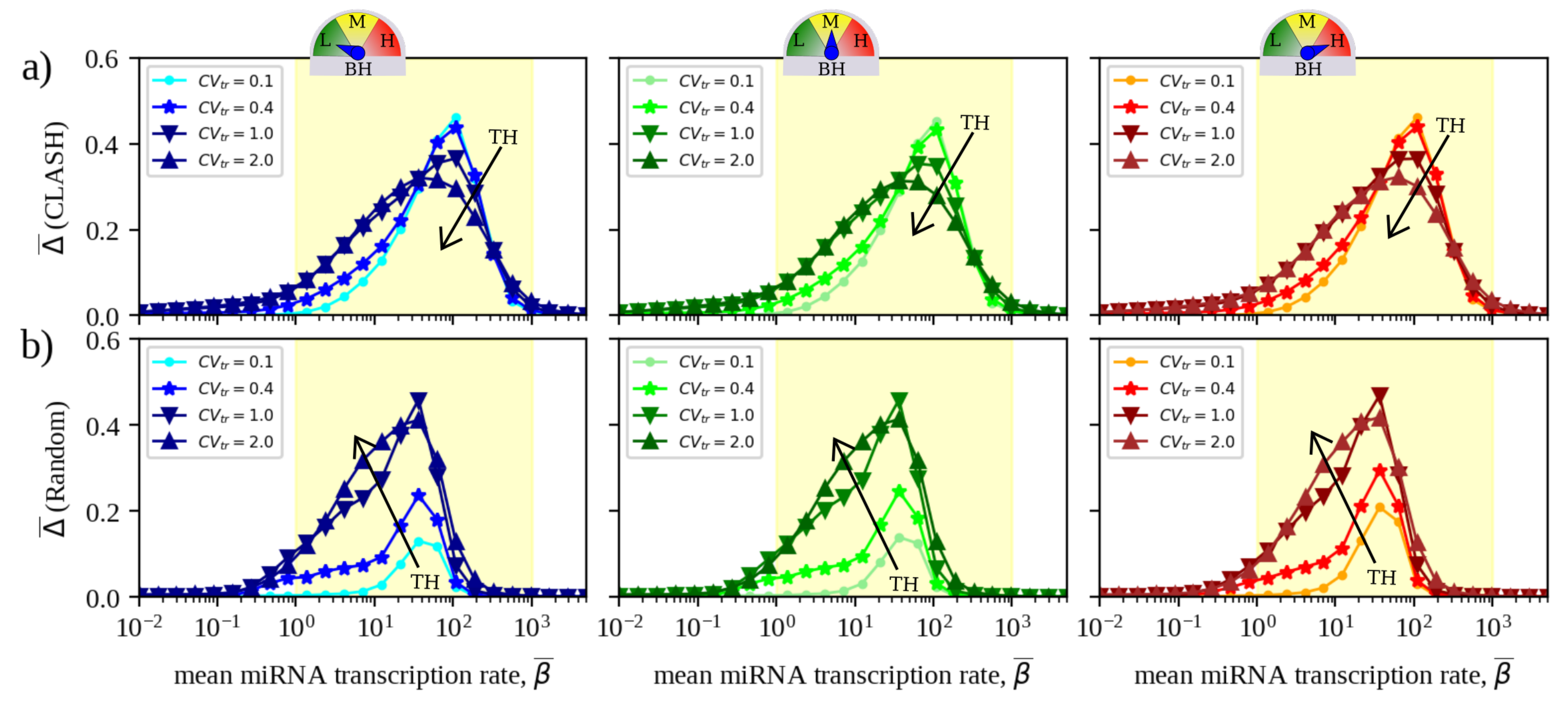}
\end{center}
\caption{ \textbf{Crosstalk asymmetry in the CLASH network and its randomized counterparts.} Profile of $\Delta$ obtained \textbf{(a)}  in the CLASH network, and \textbf{(b)} in its randomized version in the 3 scenarios considered for binding heterogeneity and for various degrees of transcriptional heterogeneity. Curves are averaged over 100 independent realizations of transcription rate profiles. Results for the random case are additionally averaged over 100 independent realizations of the randomization process. In each case, the standard error of the mean is equal to or smaller than the size of the markers.}
\label{fig:supDelta}
\end{figure*}

To measure crosstalk directionality, we define the quantity
\beq
\Delta_{ij} = \left(\frac{\chi_{ij}- \chi_{ji}}{\chi_{ij} + \chi_{ji}}\right)^2~~,
\eeq
such that $0 \leq \Delta_{ij} \leq 1$. In short, the closer $\Delta_{ij}$ is to zero (resp. one) the closer crosstalk between RNAs $i$ and $j$ is to being symmetric (resp. fully asymmetric). A global measure of asymmetry is conveniently obtained by computing the average asymmetry over all pairs of different RNAs in the network, i.e.
\beq
\Delta = \frac{2}{N(N-1)}\sum_{i,j>i}^N \Delta_{ij}~~.
\eeq
Results for this quantity are reported in Fig \ref{fig:supDelta} for both the CLASH network and its randomized variant. Crosstalk asymmetry is generically larger in the susceptible regime, more pronouncedly so in the CLASH network than in its randomized version. Notably, the asymmetry profile is roughly independent of the degree of binding heterogeneity while it is only weakly modulated by transcriptional variability in the CLASH network. As seen for the mean crosstalk intensity, this state of things suggests that the way in which crosstalk asymmetry is tuned by the mean miRNA transcription rate $\overline{\beta}$ is an inherent property of miRNA-RNA networks, that is mainly encoded in their topology. The striking difference that can be seen between the behaviour of $\Delta$ in real (CLASH) and random networks (see Fig \ref{fig:supDelta}b) supports this intuition.

\begin{table}[!b]
\label{tab:subnet}
\begin{tabular}{@{}l|c|c|c@{}}
\toprule
interaction type & nr of links & nr of RNA species & nr of miRNA species
\\
\colrule
k-mer& 3624 & 2511& 195
\\
seed-nc& 6697& 3952 & 259
\\
noseed-9nt& 2828& 2121&163 
\\
noseed& 3633&2647 &199 
\\
k-mer + seed-nc& 10749& 5262 & 312
\\
CLASH - noseed& 13674& 6043&351 
\\
CLASH (whole)& 17411 & 6943 & 383 
\\
\botrule
\end{tabular}%
\caption{Summary of the CLASH subnetwork compositions.  Each subnetwork is obtained by selecting all links associated to the same kind of interaction occurring between the miRNA-RNA couples. If the subnetwork thus obtained is disjoint, the largest connected component was selected. The term `CLASH-noseed' indicates the full CLASH network except for noseed type of interactions.}
\end{table}

\subsubsection{Results for degree-preserving randomized networks}

\begin{figure*}
\begin{center}
\includegraphics[width = 0.95\textwidth]{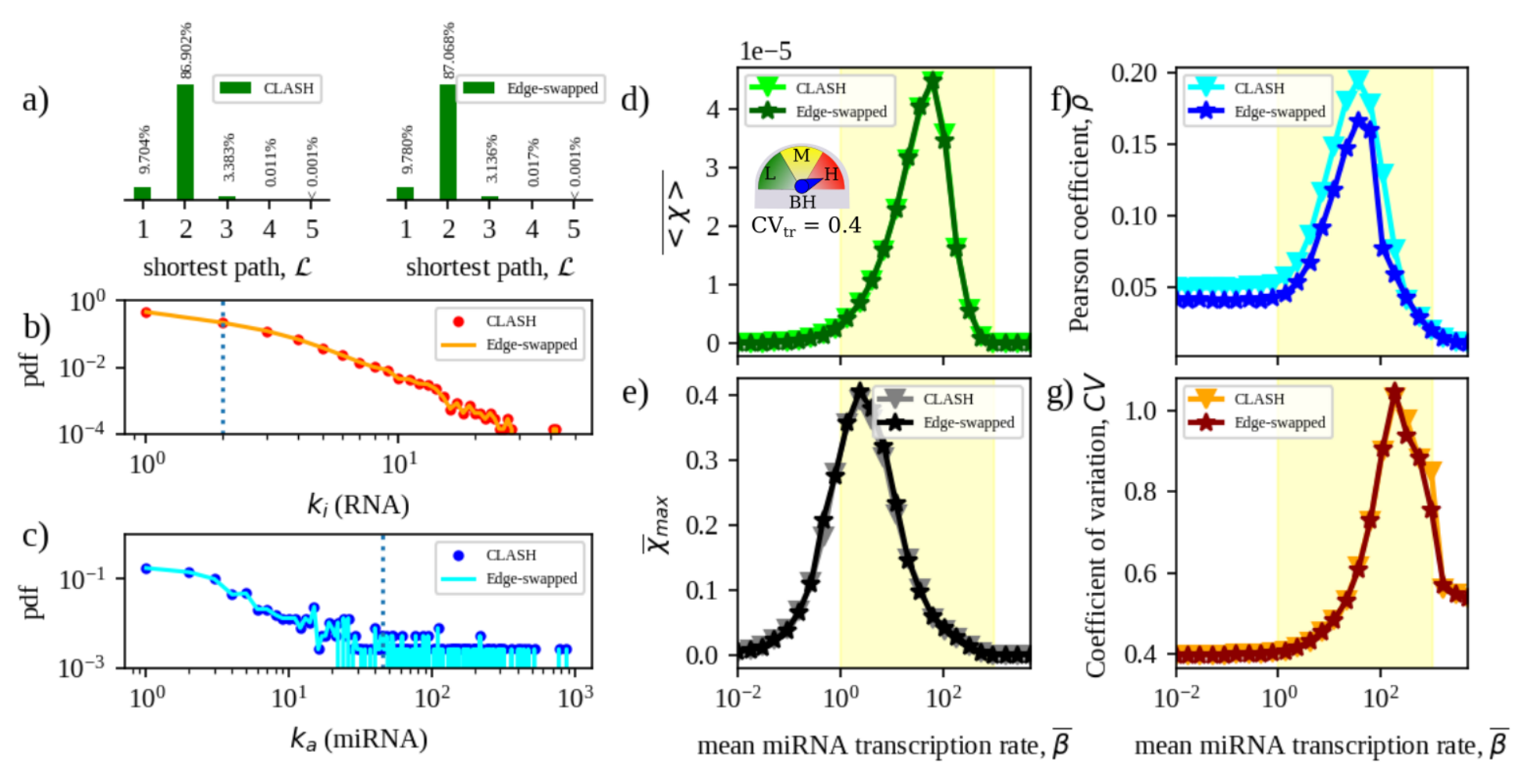}
\end{center}
\caption{\textbf{Comparison between crosstalk patterns in the CLASH network and its edge-swapped randomized versions.} \textbf{(a)} Frequency of the shortest paths for CLASH (left) and edge-swapped (right) networks. \textbf{(b, c)} Degree distributions of RNA (top) and miRNA (bottom) nodes in the CLASH and edge-swapped networks. \textbf{(d--g)} Global crosstalk descriptors for the CLASH and edge-swapped networks obtained for a degree of transcriptional heterogeneity $CV_{\mathrm{tr}}= 0.4$ and strong binding heterogeneity as a function of the mean miRNA transcription rate $\overline{\beta}$: (d) mean susceptibility; (e) mean maximum susceptibility; (f) Pearson correlation coefficient $\overline{\rho}$ between susceptibilities and local kinetic parameters; (g) Coefficient of variation of RNA levels. Averages over 100 realizations of TH in all cases except for panel (g), where 1000 realizations were taken. In each case, the standard error of the mean is equal to or smaller than the size of the markers.}
\label{fig:es}
\end{figure*}

To randomize the CLASH network while preserving the degree sequence we employ a standard edge-swapping algorithm: 
\begin{enumerate}
\item randomly select two links $\ell_{ia}$ and $\ell_{jb}$ from the miRNA-RNA network with uniform probability;
\item swap the links, obtaining new connections $\ell_{ja}$ and $\ell_{ib}$ while keeping the inverse binding affinities ($\mu^0_{ia}$ and $\mu^0_{jb}$) associated to RNAs $i$ and $j$ respectively;
\item discard the swap if it generates duplicate links or if the resulting network is not connected;
\item iterate steps 1-3 a number $n$ of times much larger than the total number of links in the network (in our case, $n=10^5$).
\end{enumerate}  
The resulting edge-swapped network has the same number of links and the same one-point statistics (i.e. the node connectivities) of the original network, while higher-order (e.g. two-node) topological correlations are lost. Numerical results were obtained by averaging over 100 independent realizations of the randomization protocol. As shown in Fig \ref{fig:es}, the structure of randomized networks differs only slightly  from that of the original CLASH network in the distribution of shortest paths between RNA species, whereas degree distributions are expectedly unchanged. In such conditions, global crosstalk descriptors are nearly identical to those obtained in the original CLASH network (panels d through g). This confirms that node degrees are the key topological determinant of the crosstalk scenario derived from the CLASH data. 

\newpage


\begin{figure*}[!h]
\begin{center}
\includegraphics[height =0.83 \textheight]{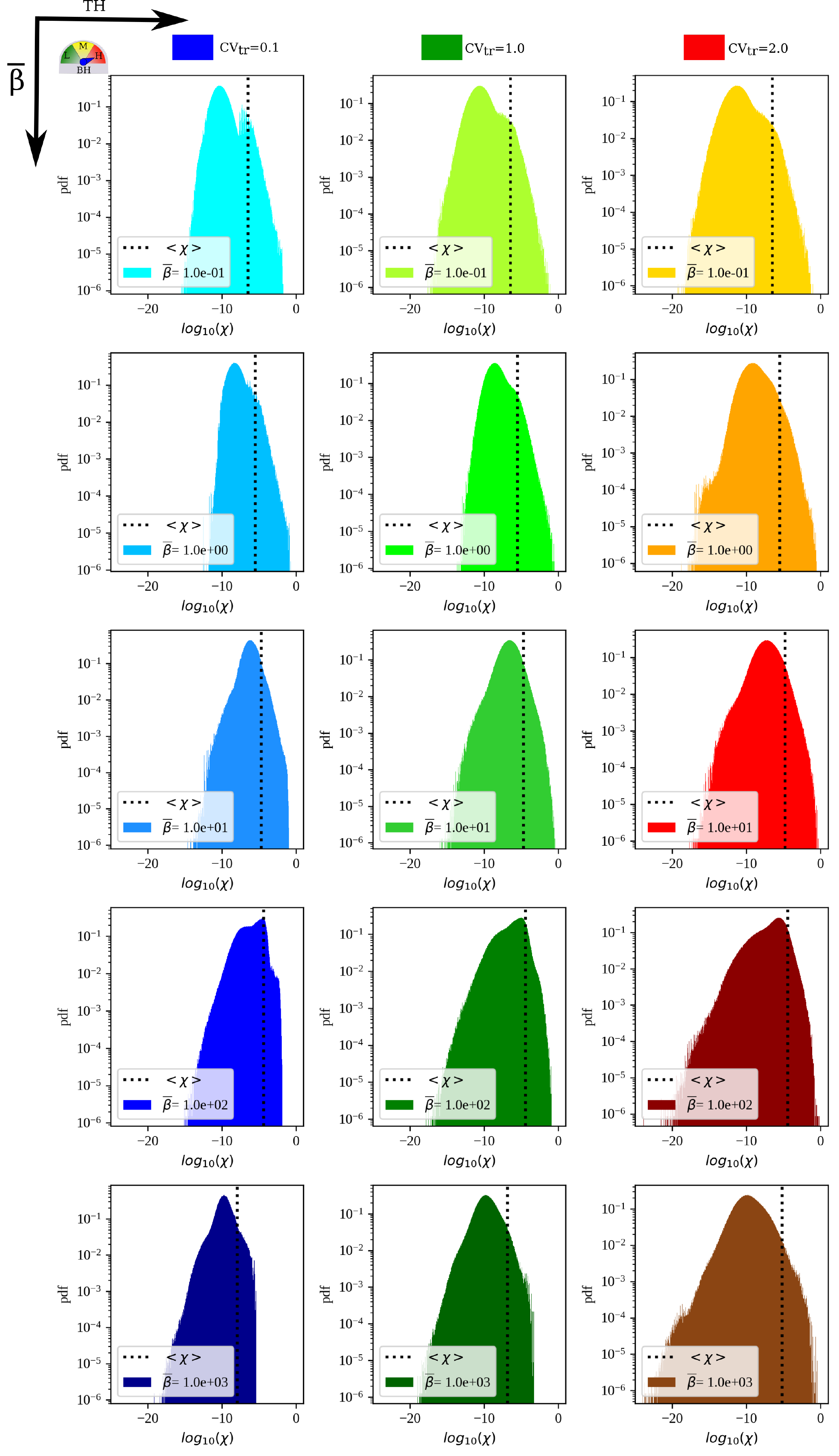}
\end{center}
\caption{\textbf{Representative susceptibility distributions.} Distributions of susceptibilities between different RNA pairs for single realizations of the CLASH interactome with different values of the mean miRNA transcription rate $\overline{\beta}$ and degrees transcriptional heterogeneity, and at fixed (high) binding heterogeneity. Note that the maximum achievable self-susceptibilities are equal to 1 (or to $\log_{10}\chi=0$).}
\end{figure*}

\newpage

\begin{figure*}[!h]
\begin{center}
\includegraphics[width=\textwidth]{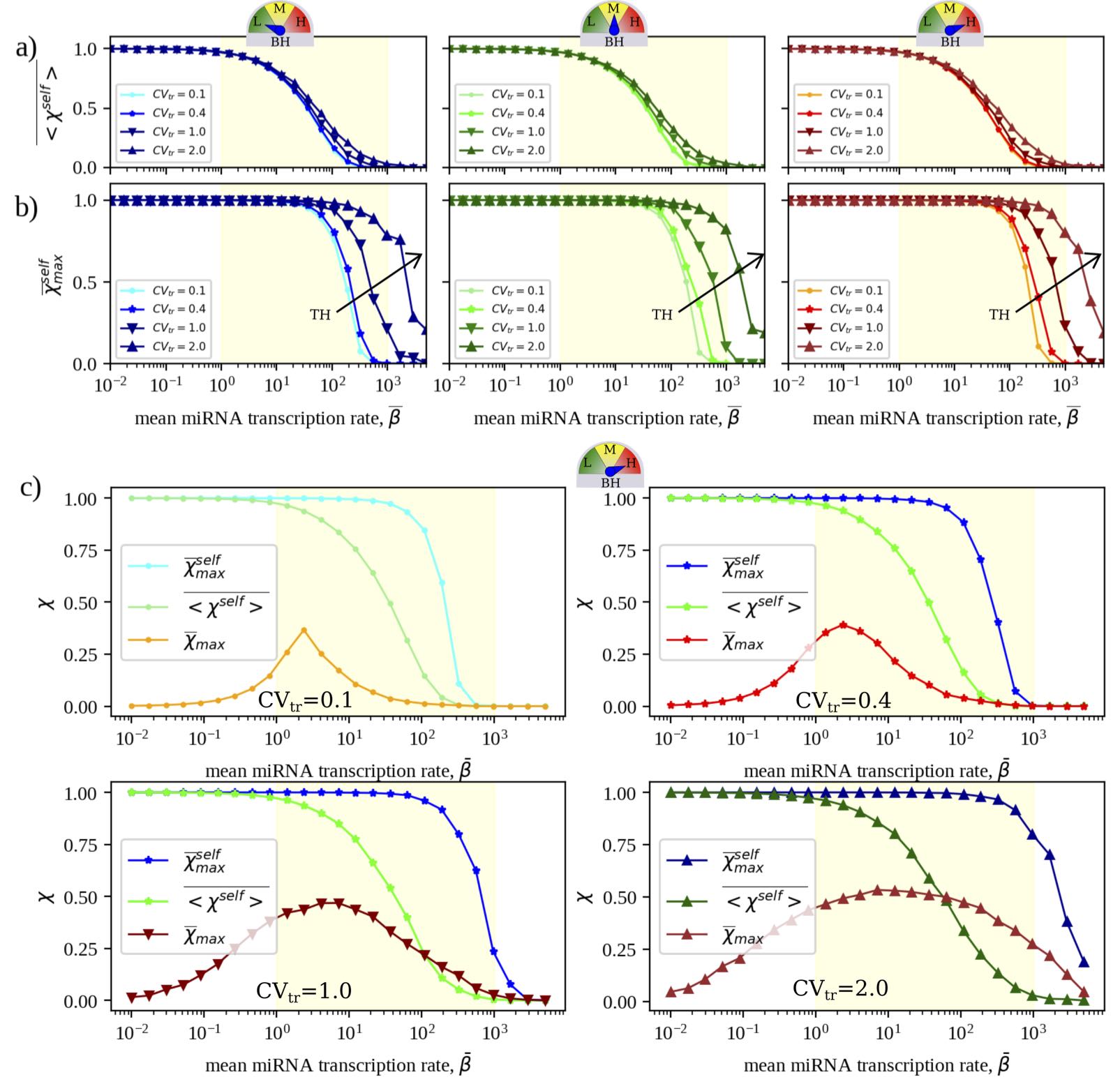}
\end{center}
\caption{\textbf{Quantitative features of RNA crosstalk in the CLASH interactome derived by solving Eq (6) (Main Text): self-susceptibilities.} \textbf{(a)} Mean self-susceptibility (averaged over RNA species and over 100 independent realizations of transcriptional heterogeneity) as a function of the mean miRNA transcription rate $\overline{\beta}$. \textbf{(b)} Mean maximal self-susceptibility (computed over all RNA species and averaged over 100 independent realizations of TH) as a function of the mean miRNA transcription rate $\overline{\beta}$. Results are shown for the 3 BH scenarios considered. Parameter values are reported in Table 1 (Main Text). The yellow shaded area qualitatively marks the region where the mean susceptibility is significantly different from zero, which coincides with the susceptible regime. In each case, the standard error of the mean is equal to or smaller than the size of the markers. The self-susceptibility is maximal when miRNA levels are low, in which case the availability of free RNA molecules increases roughly linearly with the transcription rate. As $\overline{\beta}$ increases, miRNA repression gets stronger and self-susceptibilities decrease until, at large enough $\overline{\beta}$, RNAs are fully repressed and therefore insensitive to small changes in their transcription rates. \textbf{(c)} Comparison between maximum self-susceptibility (averaged over TH realizations), mean self-susceptibility (averaged over TH realizations) and $\overline{\chi_{\max}}$ for different degrees of TH in the high BH scenario. The intensity of crosstalk between different RNAs, measured by the latter quantity, is indeed of the same order of magnitude as self-susceptibilities.}
\end{figure*}

\vspace{0cm}

\begin{figure*}[!h]
\begin{center}
\includegraphics[width = \textwidth]{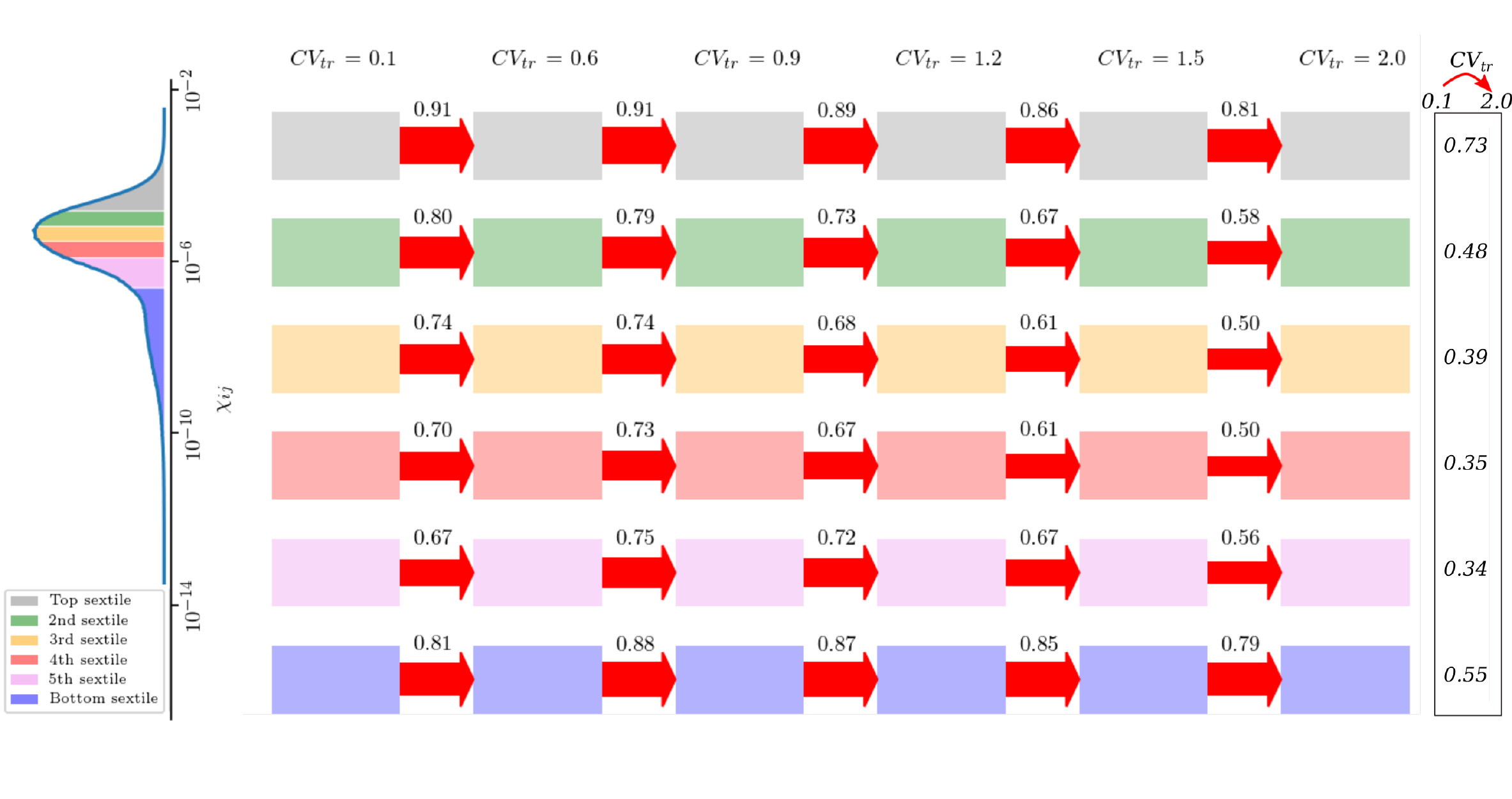}
\end{center}
\caption{\textbf{Stability of crosstalk interactions among distant RNAs in the CLASH network upon increasing degrees of transcriptional heterogeneity.} For each susceptibility sextile, we report the fraction of crosstalk interactions between distant RNAs (i.e. RNAs that do not share any miRNA regulator) that are conserved upon increasing the degree of TH. The last column reports the fraction of interactions that are conserved passing from the lowest to the highest degree of TH. Results obtained by averaging over 100 independent  realizations of transcriptional heterogeneity in each case, assuming high binding heterogeneity and mean miRNA transcription rate $\overline{\beta}=30$. Different intermediate values of $\overline{\beta}$ return qualitatively identical scenarios.}
\end{figure*}

\newpage

\begin{figure*}
\begin{center}
\includegraphics[width = 0.65\textwidth]{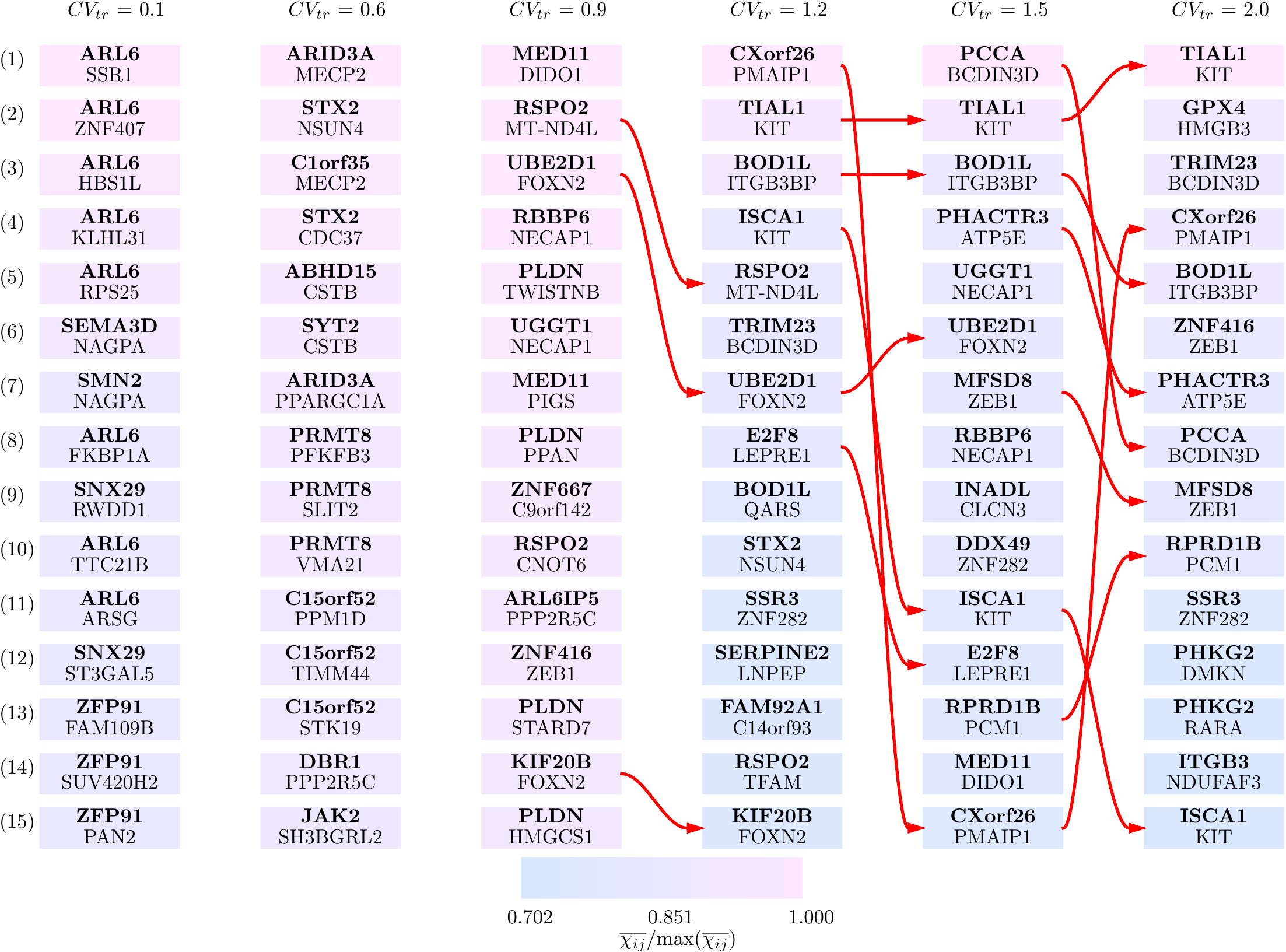}
\end{center}
\caption{\textbf{15 strongest crosstalking RNA pairs in different scenarios of transcriptional heterogeneity for the CLASH interactome.} Note that involved susceptibilities (given by color code at the bottom) are of the order of the self-susceptibility. Results were obtained by averaging over 100 independent realizations of TH for each value of $\mathrm{CV}_\mathrm{tr}$, assuming high BH and mean miRNA transcription rate $\overline{\beta}=30$. }
\end{figure*}

\begin{figure*}
\begin{center}
\includegraphics[width = 0.65\textwidth]{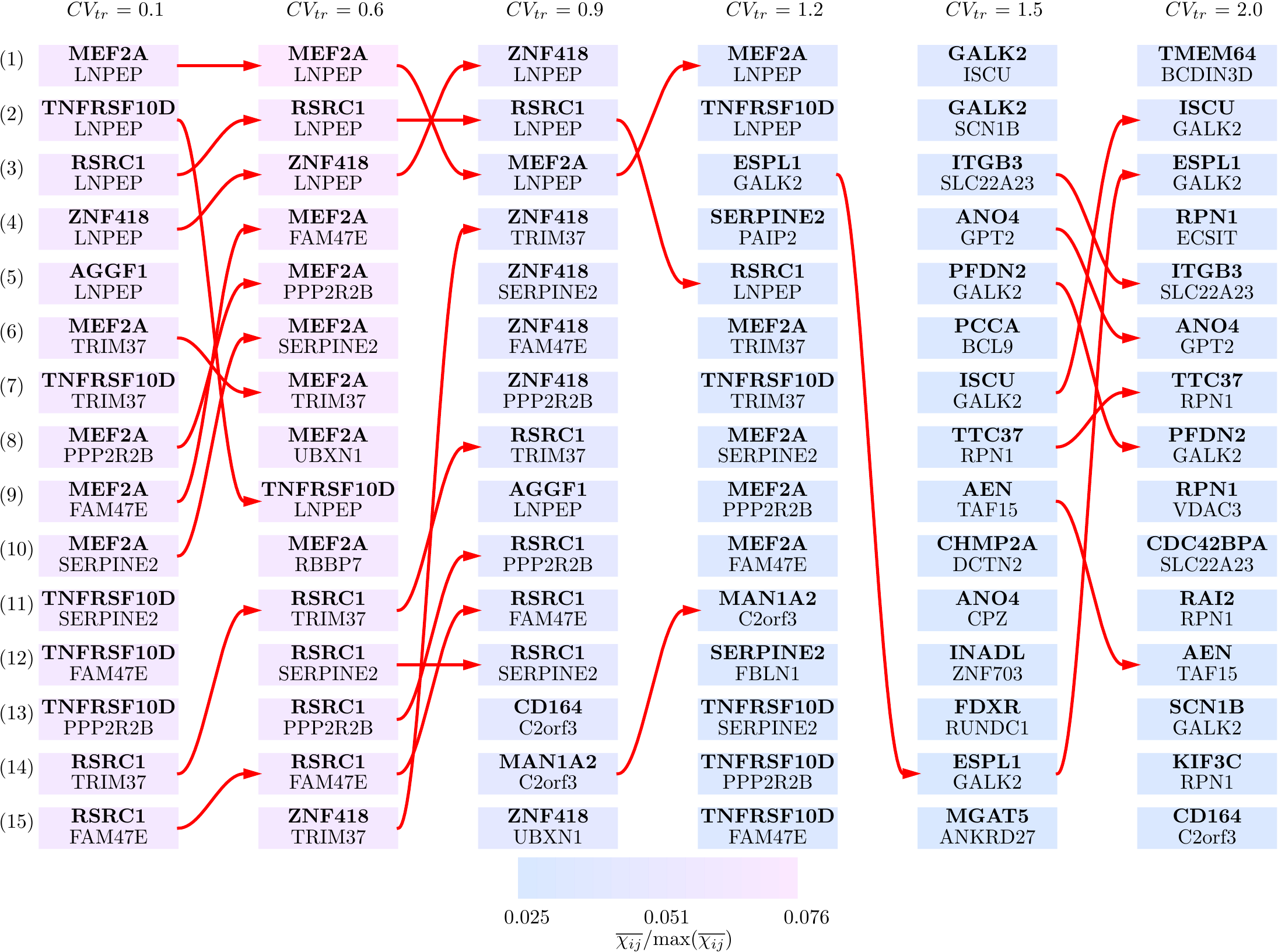}
\end{center}
\caption{\textbf{15 strongest crosstalking RNA pairs not sharing miRNA regulators in different scenarios of transcriptional heterogeneity for the CLASH interactome.} A significant degree of conservation is seen at low and high $\mathrm{CV}_{\mathrm{tr}}$. Note that involved susceptibilities (given by color code at the bottom) are roughly 2 orders of magnitude larger than the mean susceptibility (see Main Text, Fig 2). Results were obtained by averaging over 100 independent realizations of TH for each value of $\mathrm{CV}_{\mathrm{tr}}$, assuming high BH and mean miRNA transcription rate $\overline{\beta}=30$. }
\end{figure*}

\newpage

\begin{figure*}[!h]
\begin{center}
\includegraphics[width = \textwidth]{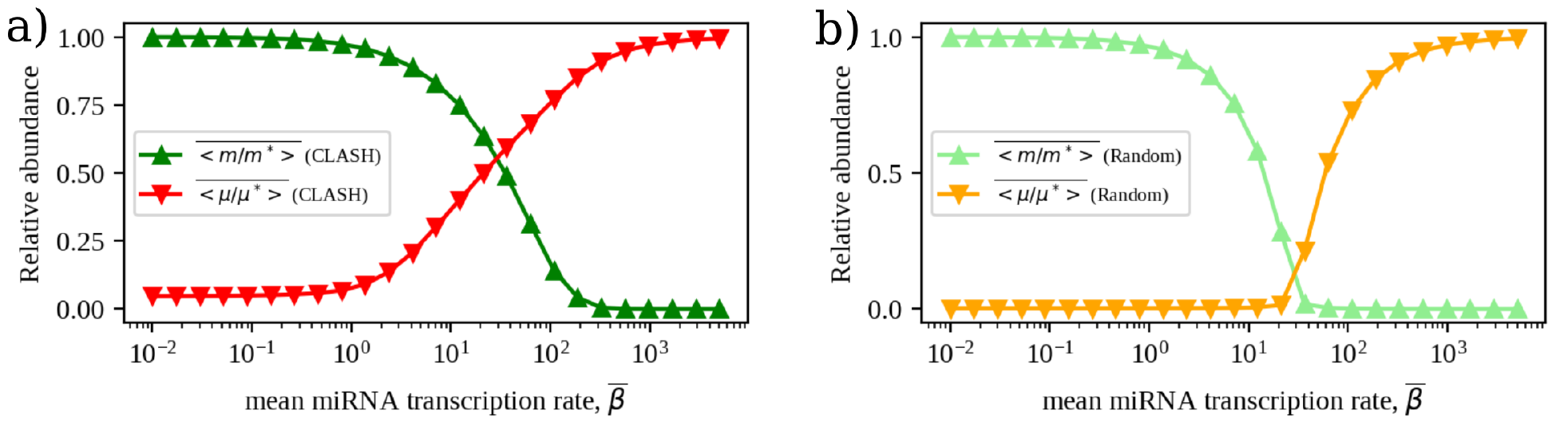}
\end{center}
\caption{ \textbf{Relative overall molecular abundances in CLASH (a) and degree-preserving randomized networks (b).} Note that the susceptible regime in the latter is narrower compared to the original CLASH network.}
\end{figure*}

\vspace{3cm}

\begin{figure*}[!h]
\begin{center}
\includegraphics[width = \textwidth]{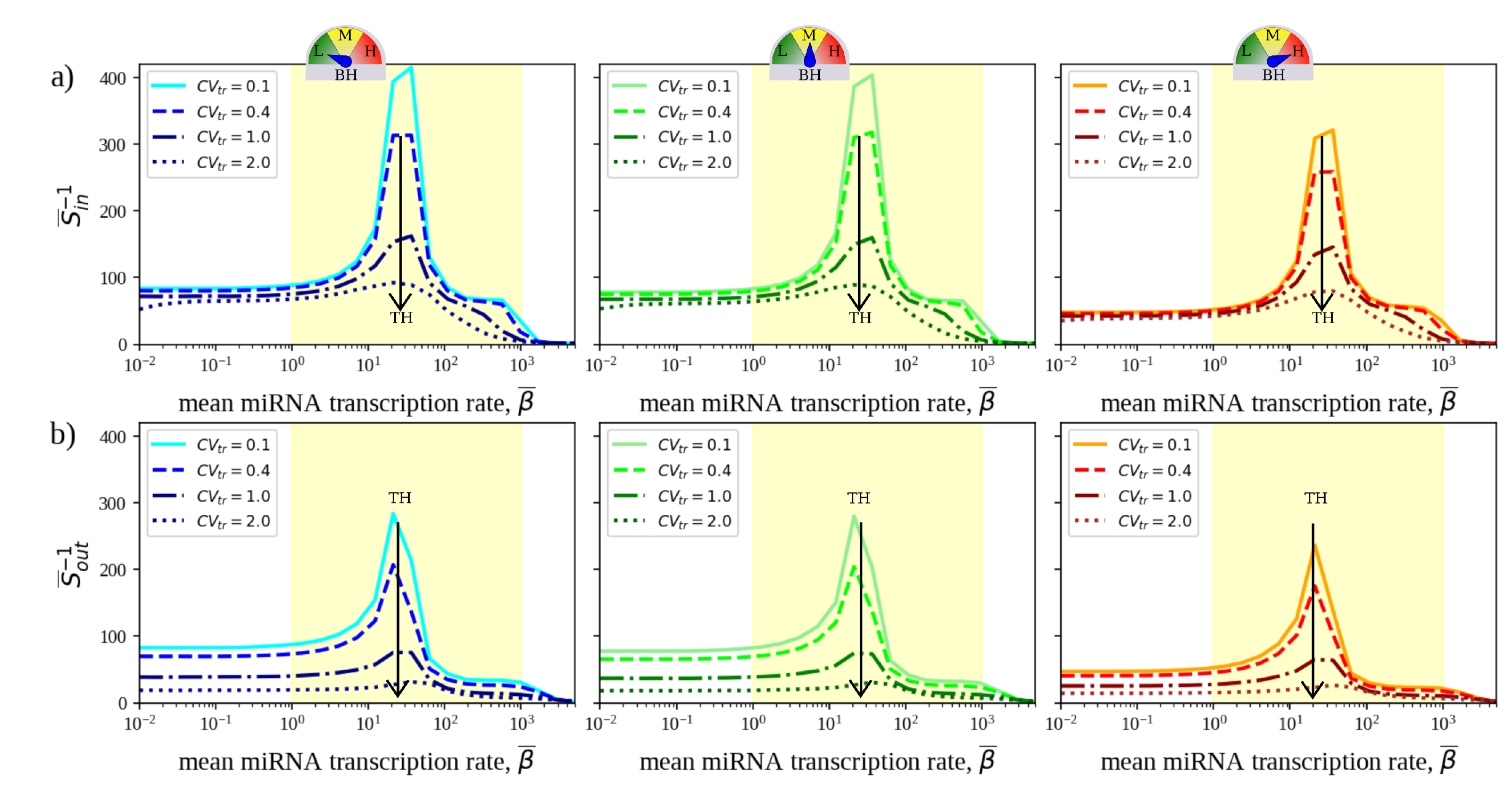}
\end{center}
\caption{\textbf{Crosstalk selectivity in randomized CLASH networks.} \textbf{(a)} Inverse of incoming and \textbf{(b)} outgoing selectivities as functions of $\overline{\beta}$ for varying degrees of TH (different curves in the same panel) and BH (reported by the 3-state gauge in different panels). Curves are averaged over 100 independent realizations of transcription rate profiles and over 100 independent realizations of the randomization process.}
\end{figure*}

\newpage

\begin{figure*}[!h]
\begin{center}
\includegraphics[width = \textwidth]{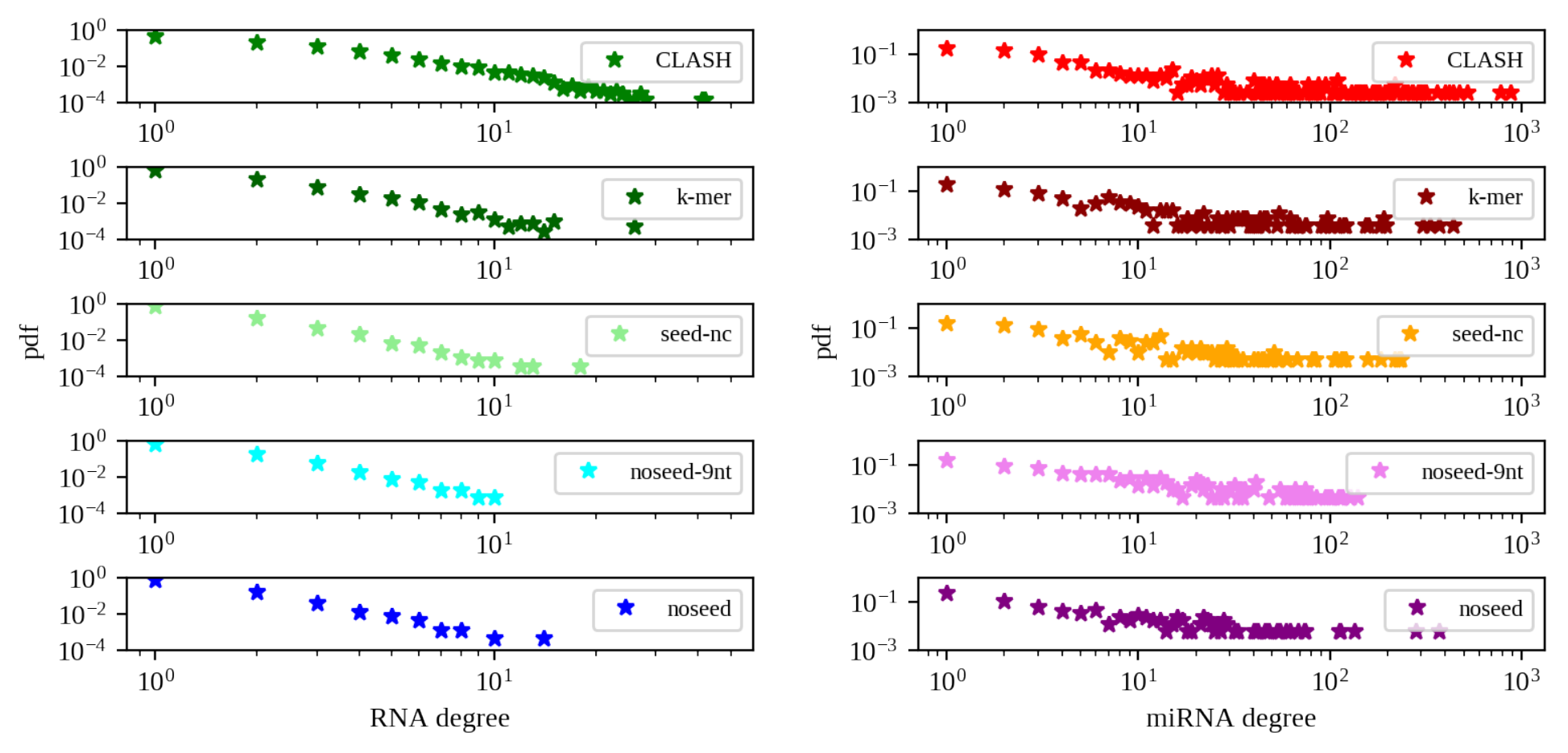}
\end{center}
\caption{\textbf{Degree distributions of CLASH subnetworks induced by individual classes of miRNA-RNA interactions.} Degree distributions for RNA nodes (left) and miRNA nodes (right) are displayed for the entire network (top panels) and for the four subnetworks defined by the interaction classes considered in this work (see Main Text, Fig 1c).}
\end{figure*}

\vspace{3cm}

\begin{figure*}[!h]
\begin{center}
\includegraphics[width = \textwidth]{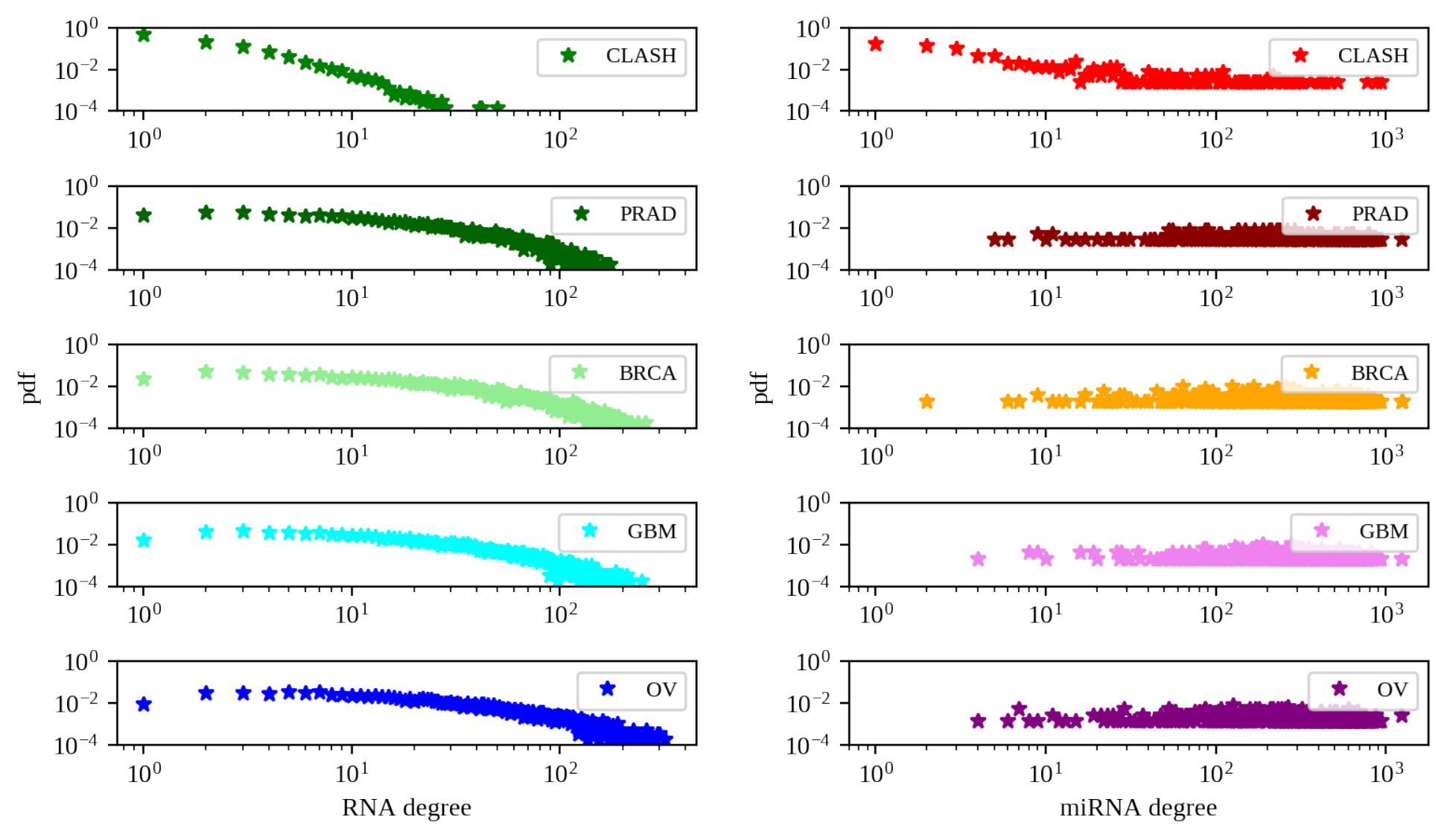}
\end{center}
\caption{\textbf{Degree distributions of the tumor specific networks reconstructed in Ref \cite{chiu}.} Degree distributions for RNA nodes (left) and miRNA nodes (right) representing the miRNA-RNA networks for prostate adenocarcinoma (PRAD), ovarian adenocarcinoma (OV), breast adenocarcinoma (BRCA) and glioblastoma (GBM) cells. Data from \cite{chiu}. Notice that the basic characteristics of degree distributions appear to be conserved across different networks. This is possibly in line with the fact that such networks present a significant context-independent component. See \cite{chiu} for a more detailed analysis.}
\end{figure*}

\newpage

\begin{figure*}[]
\begin{center}
\includegraphics[width = \textwidth]{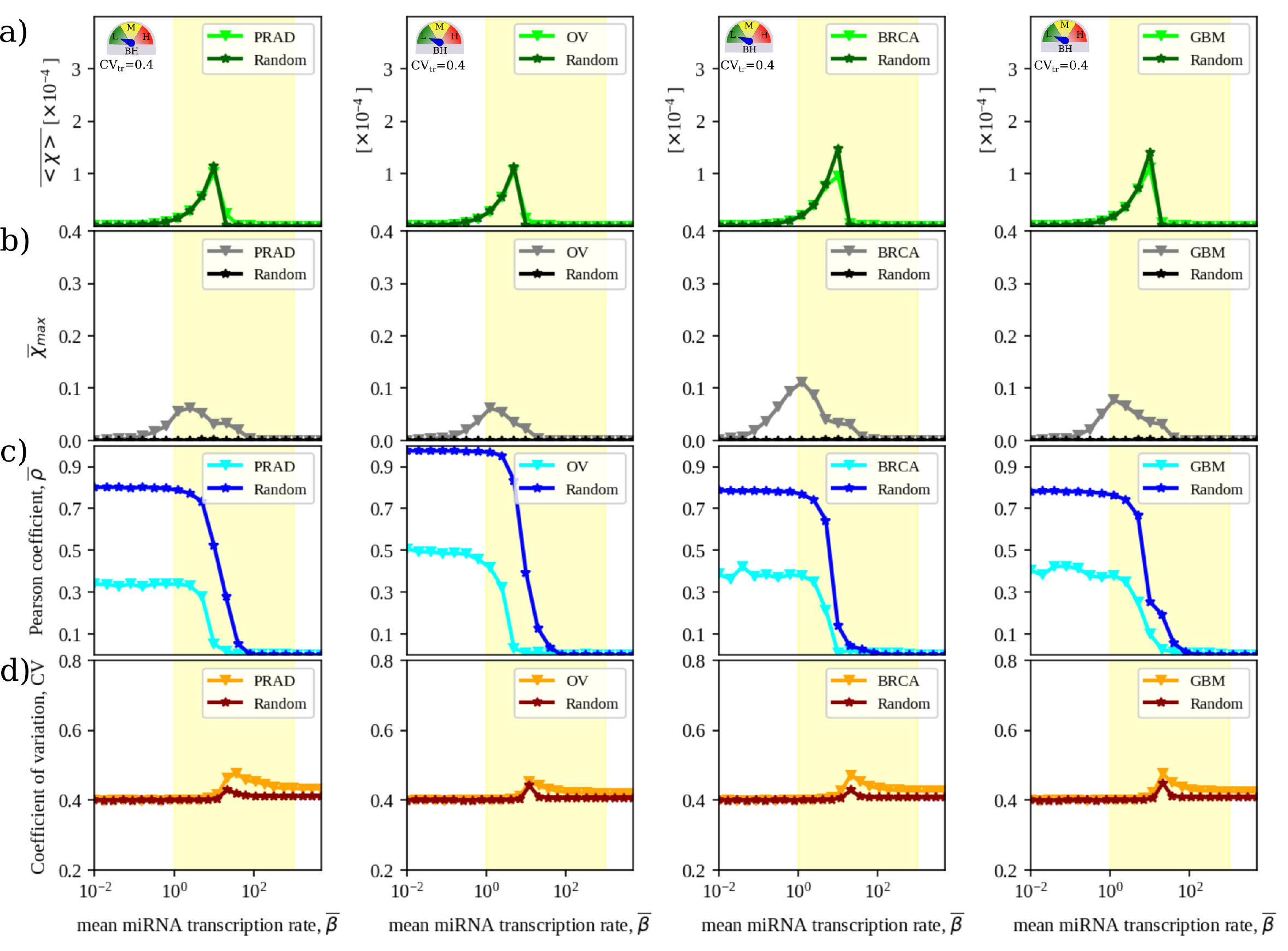}
\end{center}
\caption{ \textbf{Global RNA crosstalk descriptors derived for the tumor specific networks reconstructed in Ref \cite{chiu} as a function of the mean miRNA transcription rate $\overline{\beta}$.} Different columns represent results obtained for prostate adenocarcinoma (PRAD), ovarian adenocarcinoma (OV), breast adenocarcinoma (BRCA) and glioblastoma (GMB) cells. \textbf{(a)} Mean susceptibility. \textbf{(b)} Maximal susceptibility. \textbf{(c)} Pearson correlation coefficient between susceptibilities and local kinetic parameters. \textbf{(d)} Coefficient of variation of RNA levels. TH was set at $CV_{\mathrm{tr}} = 0.4$ and the lowest degree of binding heterogeneity was assumed. Averages were performed over 100 realizations of TH in all cases except for panels (d), where 1000 realizations were taken. Results for other cases are qualitatively similar.}
\end{figure*}

\end{document}